\documentclass[11pt]{article}

\usepackage{authblk}
\usepackage{microtype}
\usepackage{comment}
\usepackage{dcolumn}
\usepackage{xcolor}
\usepackage{youngtab}
\usepackage{ytableau}
\usepackage{tabularray}
\usepackage{booktabs}
\usepackage{placeins}
\usepackage{soul}
\usepackage{bm}

\definecolor{redd}{rgb}{0.8, 0.1,0.2}
\definecolor{navy}{rgb}{0.05, 0.23,0.75}
\definecolor{ochre}{rgb}{0.8, 0.47, 0.13}
\definecolor{navyblue}{rgb}{0.0, 0.0, 0.5}
\definecolor{oucrimsonred}{rgb}{0.6, 0.0, 0.0}
\definecolor{nicegreen}{rgb}{0.31, 0.64, 0.38}

\usepackage{geometry}
\usepackage[utf8]{inputenc}
\usepackage[absolute,overlay]{textpos}
\usepackage{slashed}
\usepackage{chngcntr}
\usepackage{hyperref}
\hypersetup{colorlinks, linkcolor={green!47!black}, citecolor={green!60!black}, urlcolor={green!55!black}}
\usepackage{bbm}
\usepackage{amsfonts}
\usepackage{amsmath,amssymb}
\usepackage{mathrsfs}
\usepackage{epsfig}
\usepackage{graphicx}
\usepackage{wrapfig}
\usepackage{url}
\usepackage[nameinlink]{cleveref}
\usepackage{float}
\usepackage{color}
\usepackage{multirow}
\usepackage{lipsum}
\usepackage{enumitem}
\usepackage{chngcntr}
\usepackage{tikz}
\usepackage{adjustbox}
\usepackage{cancel}
\usepackage[sorting = none, backend = bibtex, style=numeric-comp,maxnames=50]{biblatex}
\usepackage{lmodern}
\usepackage[T1]{fontenc}

\DeclareSymbolFont{myletters}{OML}{ztmcm}{m}{it}
\DeclareMathSymbol{\uplambda}{\mathord}{myletters}{"15}

\newcommand{\imag}{\text{i}}

\newcolumntype{x}[1]{>{\centering\arraybackslash\hspace{0.cm}}p{#1}}
\newcolumntype{L}{>{\centering\arraybackslash}m{1.5cm}}

\DeclareSymbolFont{symbolsC}{U}{pxsyc}{m}{n}

\graphicspath{{Figures_CW/}}

\newcommand{\kSSB}{k_{{\rm d}{\rm SB}}}
\newcommand{\kconf}{k_{\rm conf}}
\newcommand{\SSB}{{\rm dSB}}
\newcommand{\dSSB}{{\rm d}{\rm SB}}
\newcommand{\acrit}{\alpha^{\rm crit}_{\SSB}}
\newcommand{\aFP}{\alpha^{*}_{g}}
\newcommand{\Nfcrit}{N^{\rm crit}_{f}}

\makeatletter \gdef\@fpheader{} \makeatother

\title{Non-perturbative~news~from~the~conformal~window}

	\author[a]{\'Alvaro~Pastor-Guti\'errez}
	\affil[a]{RIKEN Center for Interdisciplinary Theoretical and Mathematical Sciences (iTHEMS), RIKEN, Wako 351-0198, Japan}
	\date{\vspace{-5ex}}
\makeindex
\begin{document}
	
\begin{textblock*}{5cm}(\dimexpr\paperwidth-6cm\relax, 1in)
	  \hspace{-2cm }RIKEN-iTHEMS-Report-26
\end{textblock*}

\maketitle  
\begin{abstract}
Dynamical symmetry breaking plays a crucial role in mass and scale generation.
In QCD-like theories, its dependence on the number of fermion flavours determines the phase structure and the transition to the conformal regime.
In this work, we compute non-perturbative corrections relevant to this phenomenom, including momentum dependencies and field invariants essential for an adequate realisation of the global symmetries, by means of  the functional renormalisation group and the generalised flow equation.
We find that the critical gauge coupling required for dynamical chiral symmetry breaking acquires a strong dependence on the number of flavours and increases sharply at $N_f^\textrm{crit}\simeq 7.30$ for $N_c=3$.
This establishes a new picture of the phase transition to the conformal window that disfavours Miransky/BKT scaling and walking regimes but supports a first-order quantum phase transition and a critical region with exotic dynamics and symmetric fermion mass gaps.
\end{abstract}
\tableofcontents
\section{Introduction}
\label{sec:introduction}
Understanding the phase structure and infrared (IR) dynamics of gauge-fermion quantum field theories (QFTs) stands as one of the central challenges in theoretical physics, and in particular for non-perturbative approaches. This endeavour is of both theoretical and phenomenological interest, as it provides insights relevant to the physical theory of Quantum Chromodynamics (QCD) and to extensions beyond the Standard Model.

QCD-like theories are known to display two defining IR phenomena: colour confinement and dynamical chiral symmetry breaking. Both appear as a consequence of the strongly coupled nature of the gauge interactions and together provide the defining features and spectrum of the theory. In this work we will focus on the aspects of dynamical symmetry breaking ($\dSSB$) and how they vary as a function of the symmetries, namely the number of colours ($N_c$) and Dirac flavours ($N_f$).

Certain limits of QCD-like theories are well understood. For few colours and flavours, one recovers physical QCD, for which a wealth of analytical and numerical studies have achieved quantitative precision in determining correlation functions and their properties. For many flavours, $N_f\geq 11N_c/2$, asymptotic freedom~\cite{Gross:1973id,Politzer:1974fr} is lost and a perturbative ultraviolet (UV) completion is no longer available. For $N_f$ just below this limit, an IR fixed point~\cite{Caswell:1974gg,Banks:1981nn} emerges, rendering the theory quantum scale invariant and defining the conformal regime of QCD-like theories, commonly referred to as the Caswell--Banks--Zaks (CBZ) window. As $N_f$ is decreased, the CBZ fixed point strengthens and eventually leads to dynamics that can break scale invariance. This intermediate region has been studied extensively, see e.g.~\cite{Sannino:2009za,Dietrich:2006cm} for a review, but remains largely unresolved. In particular, the critical number of flavours $\Nfcrit$ at which the transition between the conformal and dynamical regimes occurs, the order of this quantum phase transition, and its near-critical scaling remain open questions.

The functional renormalisation group~\cite{Wetterich:1992yh,Dupuis:2020fhh} (fRG) provides a non-perturbative framework for studying the dynamics of general gauge-fermion theories at the level of correlation functions. This Wilsonian approach has been extensively applied to physical QCD to determine both universal and non-universal properties; see  e.g.~\cite{Fu:2022gou,Fu:2019hdw,Fischer:2026vkc,Fischer:2026uni} for recent reviews. This provides a well-established foundation for disentangling the relevant dynamics and constructing systematic approximations, which we extend here beyond the physical QCD regime. The fRG has also been previously used to study the phase structure of QCD-like theories, including the boundary of the conformal window~\cite{Gies:2005as,Goertz:2024dnz} and near-conformal scaling~\cite{Braun:2010qs,Braun:2009ns,Goertz:2024dnz}. In this work, we improve on these studies by including previously unconsidered non-perturbative corrections.

The relevant off-shell dynamics underlying $\dSSB$ are encoded in higher-order fermionic interactions. To access these, we employ scale-dependent field transformations~\cite{Gies:2001nw,Gies:2002hq,Pawlowski:2005xe,Floerchinger:2009uf} that map the fermionic sector onto a bosonised Yukawa formulation. Within the generalised flow equation~\cite{Pawlowski:2005xe}, this procedure can be implemented exactly without introducing any double counting. The resulting formulation improves upon previous treatments in two respects: it retains the momentum dependence of the bosonised channel beyond the point-like approximation and includes higher-order chiral invariants required to realise the full flavour symmetry. Both contributions introduce non-trivial leading $N_f$ dependencies into the quantum corrections and therefore affect $\dSSB$ in the many-flavour regime.

Integrating the flow equations we can determine the critical gauge coupling $\acrit$ required to trigger $\dSSB$.  We find that the newly included corrections induce an important $N_f$ dependence of $\acrit$, with increasingly strong gauge interactions required to generate a chiral condensate. This qualitatively different behaviour implies that the transition between the dynamical and conformal phases~\cite{Miransky:1996pd} occurs in a strongly coupled regime, where confinement or other non-perturbative mechanisms may become relevant. It also challenges the standard picture of walking dynamics and Miransky/BKT scaling~\cite{Braun:2010qs,Braun:2009ns,Goertz:2024dnz,Kaplan:2009kr} in the near-conformal region. In particular, our results indicate a first-order quantum phase transition and suggest a critical region in which novel dynamics, such as confinement without $\dSSB$ or symmetric fermion mass gaps, may emerge.

This work is organised as follows. In \cref{sec:effectiveaction}, we introduce the effective action, the generalised flow equation and the bosonisation procedure. In \cref{sec:critical}, we integrate the flow equations and determine $\acrit$, analysing how the newly included corrections modify its $N_f$ dependence and the resulting phase structure. In \cref{sec:conformalPT}, we discuss the implications for near-conformal scaling and the nature of the quantum phase transition, including the possible emergence of confinement without $\dSSB$ and symmetric fermion gaps. Finally, in \cref{sec:conclusions}, we summarise our findings and conclude.

\section{Effective action and $\dSSB$ in QCD-like theories}
\label{sec:effectiveaction}
The functional renormalisation group (fRG) provides a non-perturbative framework for studying the IR dynamics of gauge-fermion theories. One of its central objects is the scale-dependent effective action $\Gamma_k[\Phi]$~\cite{Wetterich:1989xg}, which incorporates quantum fluctuations between a reference UV scale ($\Lambda_{\rm UV}$) and an IR scale $k$. IR modes with momenta $p<k$ are suppressed by a regulator function $R_k$ that is quadratic in the fields. The resulting effective average action interpolates between the microscopic action at $k=\Lambda_{\rm UV}$ and the full quantum effective action at $k\to0$, where the regulator vanishes. Its scale dependence is governed by the flow equation introduced in~\cite{Wetterich:1992yh}, which accounts for quantum fluctuations in a non-perturbative fashion. Although the flow equation is exact, solving it for a general effective action requires truncations and therefore introduces approximations.

In gauge-fermion QFTs, the dynamics are encoded in the 1PI correlation functions of the effective average action,
\begin{align}\label{eq:fulleffectiveaction}
	\Gamma_{k}[A_\mu,\bar c, c, \bar\psi, \psi]
	= \Gamma_{{\rm gauge},\,k}[A_\mu,\bar c, c]
	+ \Gamma_{{\rm gauge\text{-}fermion},k}[A_\mu, \bar\psi, \psi]
	+ \Gamma_{{\rm fermion},k} [\bar\psi, \psi]\,.
\end{align}
The three terms separate the pure gauge, gauge-fermion, and fermionic sectors. The pure gauge sector contains the dynamics responsible for colour confinement, which manifests itself through the generation of a mass gap in the gauge-field propagator, see e.g.~\cite{Alkofer:2000wg} for a review. The gauge-fermion sector transmits the strength of the pure gauge dynamics to the fermionic sector, which contains the self-interactions responsible for fermion condensation, $\langle\bar\psi\psi\rangle$, and the associated breaking of chiral symmetry and scale invariance. These are therefore driven by the gauge coupling $\alpha_g=g^2/(4\pi)$ which constitutes the single marginally relevant parameter in the microscopic theory and sets the scale for the RG evolution.

The symmetries of the matter sector constrain the allowed interactions. In the chiral limit, fermions in the fundamental representation of $SU(N_c)$ possess the global flavour symmetry
\begin{align}
	{ U}(N_f)_L\times{ U}(N_f)_R
	\simeq SU(N_f)_L\times SU(N_f)_R\times U(1)_V\times U(1)_A\,,
	\label{eq:chiralsymmetry}
\end{align}
where the right-hand side neglects discrete quotient groups that are irrelevant for the present discussion. The axial $U(1)_A$ symmetry is explicitly broken by the ABJ anomaly~\cite{Adler:1969gk,Bell:1969ts}, sourced by topological gauge configurations. The lowest-order fermionic interaction this generates is the 't Hooft determinant~\cite{tHooft:1976snw}, which contains $2N_f$ fermion fields. Its canonical dimension is $3N_f$ in four dimensions, making it increasingly irrelevant as $N_f$ grows. Consequently, in the large-$N_f$ regime considered here, the axial anomaly is therefore infinitesimally broken and parametrically suppressed, and ${\rm U}(N_f)_L\times{\rm U}(N_f)_R$ provides a good approximate symmetry. 

The symmetries in \eqref{eq:chiralsymmetry} constrain the operators generated by the RG flow. At the four-fermion level, the fermionic part of the effective action reads
\begin{align}
	\Gamma_{{\rm fermion},\,k}[\bar \psi,\psi]
	= -\int_x Z_\psi^2\bigg\{
	\lambda_\textrm{\tiny{SP}} {\cal T}_{\left({\rm S-P}\right)}
	+\lambda_{+} {\cal T}_{\left({\rm V+A}\right)}
	+ \lambda_{-} {\cal T}_{\left({\rm V-A}\right)}
	+ \lambda_{\textrm{\tiny{adj}}} {\cal T}_{\left({\rm V-A}\right)^{\rm adj}}
	\bigg\}+\dots\,,
	\label{eq:eff4Fermi}
\end{align}
where $Z_\psi^2$ denotes the fermion wave-function renormalisation and $\lambda_\textrm{\tiny{SP}},\lambda_\pm,\lambda_{\textrm{\tiny{adj}}}$ are the couplings of the tensor structures ${\cal T}_i$, which form the Fierz-complete basis given in \eqref{eq:4FermiTensors}. The ellipsis accounts for other many-fermion interactions.

The fermionic operators in \eqref{eq:eff4Fermi} are generated by the gauge dynamics through gauge-fermion interactions in the form of box diagrams, and subsequently feed back into their respective RG flow~\cite{Miransky:1979ks,Braun:2011pp,Kubota:1999jf,Kondo:1991yk,Kondo:1993jq,Kondo:1992sq,Miransky:1988gk,Goertz:2024dnz}. Although these operators are canonically irrelevant, their RG evolution is driven by the relevant gauge coupling and as the gauge interaction grows towards the IR, the four-fermion couplings increase and eventually lead to a divergence. Through a Hubbard--Stratonovich transformation~\cite{HubbardPhysRevLett.3.77,Stratonovich}, this singularity is understood as a flat potential in the associated bosonic mode and the signalling of the formation of a fermion condensate which gives mass to the fermions and consequently breaks~\cite{Vafa:1983tf}
\begin{align}
	{ U}(N_f)_L\times{ U}(N_f)_R
	\longrightarrow { U}(N_f)_V
	\cong { SU}(N_f)_V\times{ U}(1)_V\,.
	\label{eq:SSBpattern}
\end{align}
This RG description of $\dSSB$ \cite{Nambu:1961fr} dates back to~\cite{Gross:1974jv,Miransky:1979ks,Rosenstein:1990nm,Zinn-Justin:1991ksq} and has since been developed extensively. In particular, the divergence can be understood as the disappearance of an IR-attractive partial fixed point in the flow of the four-fermion couplings via a merger~\cite{Fomin:1983kyk,Miransky:1989qc,Aoki:1999dv}.

Furthermore, in the broken phase, the scalar-pseudoscalar tensor structure decomposes into four channels with distinct spin-flavour quantum numbers,
\begin{align}
	\lambda_\textrm{\tiny{SP}} {\cal T}_{\left({\rm S-P}\right)}
	\longrightarrow
	\lambda_{\sigma}\, {\cal T}_{\sigma}
	+ \lambda_{\pi}\, {\cal T}_{\pi}
	+ \lambda_{\eta}\, {\cal T}_{\eta}
	+ \lambda_{a}\, {\cal T}_{a}\,.
	\label{eq:SPchannelbroken}
\end{align}
These channels correspond to the mesonic modes that describe the low-energy dynamics. After $\dSSB$, the scalar and pseudoscalar modes become non-degenerate and the anomalous $U(1)_A$ breaking  further splits the $(\mathcal{T}_\sigma,\mathcal{T}_\pi)$ and $(\mathcal{T}_\eta,\mathcal{T}_a)$ sectors, leaving the $\pi$ ($\eta$) modes as the corresponding (pseudo-)Goldstone bosons.

\subsection{Scale-dependent field transformations}
\label{sec:dynamical bosonisation}
The bosonised formulation provides a natural description of $\dSSB$, connects directly to the IR degrees of freedom in the broken phase, and offers several computational advantages. However, changing the description at a fixed scale can introduce double-counting problems~\cite{Gies:2006wv}. These can be avoided by performing field redefinitions continuously along the RG flow~\cite{Gies:2001nw,Pawlowski:2005xe,Floerchinger:2009uf}. This is implemented in a mathematically exact way through the generalised flow equation~\cite{Pawlowski:2005xe},
\begin{align}
	\left( \partial_t + \int_x \dot\Phi \frac{\delta}{\delta \Phi} \right) \Gamma_k [\Phi ] = \frac{1}{2} \textrm{Tr}\left[ \frac{1}{\Gamma^{(\Phi\Phi)}_k[\Phi] +R_k }\,\left(  \partial_t +2 \frac{\delta \dot\Phi}{\delta\Phi}  \right)  R_k  \right]\,.
	\label{eq:GenfRG}
\end{align}
Here, $\partial_t=k\partial_k$ and we use the shorthand notation for the functional derivatives
\begin{align}
	\Gamma_k^{(\Phi_{i_1}\cdots \Phi_{i_n})}\left[\Phi\right]=\frac{\delta^{n}}{\delta \Phi_{i_1}\cdots\delta\Phi_{i_n}}\Gamma_k[\Phi]\,.
	\label{eq:npointfunct}
\end{align}
The quantity $\dot\Phi[\Phi]$, written in superfield notation, parametrises the scale-dependent field transformation and can be chosen freely.

To access the deep IR and the chirally broken phase of gauge-fermion theories, we introduce the auxiliary bosonic fields
\begin{align}
	&\phi^i= (\sigma,\,{\pi}^a,\,{a}^a,\,\eta)&&{\rm with}& & {\pi}^a=(\pi_1,\,\ldots,\,\pi_{N_f^2-1})& &{\rm and}& &{a}^a=(a_1,\,\ldots,\,a_{N_f^2-1})\,,\notag&
\end{align}
and arrange them into the matrix
\begin{align}
	\Sigma = T^0_f(\sigma + \imag\eta) + T^a_f(a^a + \imag\pi^a)\,,
	\label{eq:mesonmatrix}
\end{align}
which transforms as $\Sigma\to V_L\Sigma V_R^\dagger$ under the symmetry in \eqref{eq:chiralsymmetry}. Here, $T^0_f=\mathbbm{1}_{N_f}/\sqrt{2N_f}$ is the normalised flavour-singlet generator, while $T^a_f$ ($a=1,\ldots,N_f^2-1$) are the generators of $SU(N_f)$ satisfying $\mathrm{Tr}[T^a_f T^b_f]=\delta_f^{ab}/2$. At each scale, these composite fields are aligned with the corresponding fermion bilinears according to
\begin{align}
	\langle\partial_t\hat\phi^i_k\rangle = \dot{\mathcal{A}}_{i,\,k}\,\bar\psi\,\mathbbm{T}_i\psi\,,
	\label{eq:dynbos_fields}
\end{align}
where the mean fields $\phi^i=\langle\hat\phi^i_k\rangle$ are kept $k$-independent and the tensors $\mathbbm{T}_i$ encode the spin and flavour structure of the corresponding channels, as defined in \eqref{eq:mathbbmT}. The scale dependence of the field transformation is encoded in $\dot{\mathcal{A}}_{i,k}$, which allows the four-fermion interactions to be rewritten exactly in bosonised form. At the level of the dimensionless couplings $\bar \lambda_i=\lambda_i k^2$, their flow reads
\begin{align}\label{eq:4Fermiflow}
	\partial_t \bar \lambda_i - (2+ 2\eta_\psi)\bar \lambda_i -h_i \, \dot{\mathcal{A}}_{i}= {\rm Tr} \left[ {\mathbbm P }^{( \bar \psi \psi \bar \psi \psi )}_i \partial_t \Gamma_k^{( \bar \psi \psi \bar \psi \psi )}\right]= \overline{\rm Flow}^{( \bar \psi \psi \bar \psi \psi )}_i\,.
\end{align}
The second term on the left-hand side accounts for the canonical scaling and the anomalous dimension of the fermion fields, while the third originates from the scale dependence of the field transformation. On the right-hand side, the four-fermion flow is projected onto the chosen channel using ${\mathbbm P }^{( \bar \psi \psi \bar \psi \psi )}_i$. We can therefore choose
\begin{align}
	\label{eq:dynbos_cond}
	&\bar \lambda_{i}\equiv0\,, \qquad \forall \,\, k \,& & {\rm setting }& & 
	\dot{\mathcal{A}}_{i}	\equiv -\overline{\rm Flow}^{( \bar \psi \psi \bar \psi \psi )}_i /h_{i}\,,
\end{align}
which completely bosonises the selected four-fermion channel and prevents double counting. 

The matter sector in \eqref{eq:fulleffectiveaction} then becomes
\begin{align}
	\Gamma_{\textrm{fermion},\,k}[\bar\psi, \psi]\longrightarrow
	\Gamma_{\textrm{fermion},\,k}[\bar\psi, \psi]_{{\cal T}_{\textrm{(S-P)}}=0}
	+ \Gamma_{{\rm boson},\,k}[\bar\psi, \psi,\phi]\,,
	\label{eq:FinalGmat}
\end{align}
where the first term contains the unbosonised ${\cal T}_{({\rm V}\pm{\rm A})}$ and ${\cal T}_{({\rm V-A})^{\rm adj}}$ channels, and the second contains the dominant scalar-pseudoscalar sector in  the Yukawa form
\begin{align}	\label{eq:effbosonised}
	\Gamma_{{\rm boson},\,k}[\bar\psi, \psi,\phi] &=\int_x \biggl\{
	Z_{\phi_i}\,\mathrm{Tr}\!\left[(\partial_\mu\Sigma^\dagger)(\partial^\mu\Sigma)\right]
	+ V(\Sigma^\dagger\Sigma)+  Z^{1/2}_{\phi_i} Z_\psi\,h_i \,\bar{\psi}\, \mathbbm{T}_i \,\phi_i \psi\biggl\}\,.
\end{align}
The first term describes the propagation of the composite fields with their wave-function renormalisation $Z_{\phi_i}$ and $V(\Sigma^\dagger\Sigma)$ stands for the full mesonic potential. The last term is the Yukawa interaction between the fermions and composite bosons.

The equivalence between the bosonised and four-fermion descriptions is evident when evaluating \eqref{eq:effbosonised} on the scalar equations of motion,
\begin{align}
	\phi^i_{\textrm{\tiny{EoM}}}(p) = -\frac{Z_\psi\, h_i}{	Z_{\phi_i}^{1/2} \left(p^2+m_{\phi_i}^2\right)} \int_q \bar \psi(q) \,\mathbbm{T}_i \, \psi(p-q)\,,
	\label{eq:EoMphi}
\end{align}
which substituting this  back into the bosonised action reproduces the original four-fermion interaction, with
\begin{align}\label{eq:lambdaEoMphi}
	\lambda_i (p)\equiv\frac{1}{2} \frac{h_i^2}{p^2 +m_{\phi_i}^2}\,.
\end{align}
This relation holds at every RG scale $k$ and also makes explicit how bosonisation includes the momentum dependence of the four-fermion coupling naturally going beyond the point-like limit.

So far, we have kept all (S--P) channels independent, as required to describe the broken phase, where the different modes become non-degenerate. However, from here on we now specialise to the symmetric regime where chiral symmetry identifies a single $Z_\phi$ and $h$ for all scalar-pseudoscalar modes. 
Although the general formulation so far used sets the stage for a borken phase investigation, here instead we want to focus on the advantages that bosonisation provides in the symmetric phase. First, the bosonic dispersion relation in \eqref{eq:effbosonised} and \eqref{eq:lambdaEoMphi} allow us to retain the momentum dependence of the four-fermion interaction in the selected channel, going beyond the local point-like approximation~\cite{Jaeckel:2003uz,Gies:2006wv,Braun:2011pp}. Second, the bosonic potential contains mesonic self-interactions, which correspond to fermionic interactions beyond the four-fermion level. Together, these features provide a truncation capable of capturing non-perturbative corrections which we will show to be important.

\subsection{Chiral invariants and bosonic potential}
\label{sec:potential}
In the classical limit, the bosonic potential is invariant under $U(N_f)_L\times U(N_f)_R$. Its self-interactions can therefore be expressed in terms of the independent field invariants ${\rm inv}_n=\mathrm{Tr}[(\Sigma^\dagger\Sigma)^n]$, with $n=1,\ldots,N_f$~\cite{Pisarski:1983ms,Jungnickel:1995fp}. The two lowest-order invariants can be conveniently written as
\begin{align}	\label{eq:rhotau} 
	&	\rho = \mathrm{Tr}\!\left[\Sigma^\dagger\Sigma\right]
	= \frac{1}{2}\left(\sigma^2+\eta^2+\boldsymbol{a}^2+\boldsymbol{\pi}^2\right)& {\rm and } \quad& \tau = \mathrm{Tr}\!\left[\,\left(\Sigma^\dagger\Sigma - \frac{\rho}{N_f}\mathbbm{1}_{N_f}\right)^{\!2}\,\right]\,,&
\end{align}
with canonical dimensions $[\rho]=k^2$ and $[\tau]=k^4$ which map, respectively, onto four- and eight-fermion interactions. Higher-order invariants ${\rm inv}_{n\geq3}$ generate operators with canonical dimension $\geq8$ and correspond to subleading sixteen-fermion interactions or higher. We neglect these contributions in this work.

The $\tau$ invariant is \textit{essential} for realising the correct flavour symmetry in~\eqref{eq:chiralsymmetry}. Setting $\tau=0$ enlarges the classical symmetry of the potential to that of  $SO(2N_f^2)$~\cite{Pisarski:1983ms,Jungnickel:1995fp,Resch:2017vjs,Fejos:2024bgl,Fejos:2023lvw}, effectively describing a different theory. Its role therefore goes beyond adding quartic fluctuations as it is required to reproduce the correct physical spectrum. In the broken phase, it controls the mass splitting $m_a-m_\pi=4\rho\partial_\tau V/N_f\,,$ and hence affects the number of critical modes. In the symmetric regime, $\tau$ also has a sizeable impact as introduces $N_f$ dependencies in the corrections to the small-field curvature of the potential.

We resolve the field dependence of the potential through a polynomial expansion 
\begin{align}
	u(\bar\rho,\bar\tau) = \sum_{n=0}^{N_{\rm max}} \frac{\lambda_n}{n!}\,\bar\rho^n
	+ \sum_{m=0}^{M_{\rm max}} \frac{\lambda^{(\tau)}_m}{m!}\,\bar\tau^m
	+ \sum_{l=0}^{L_{\rm max}} \frac{\lambda^{(\rho\tau)}_l}{l!}\,\bar\rho^l\,\bar\tau\,,
	\label{eq:potential}
\end{align}
where $u(\bar\rho,\bar\tau)=V(\rho,\tau)/k^4$ is the dimensionless effective potential, expressed in terms of the renormalised dimensionless invariants $\bar\rho=Z_\phi\,\rho/k^2$ and $\bar\tau=Z_\phi^2\,\tau/k^4$. The zeroth-order coefficients vanish, but are retained here to study the convergence of the truncation as a function of $N_{\rm max}$, $M_{\rm max}$, and $L_{\rm max}$. This potential maps to fermionic interactions of orders $(\bar\psi\psi)^{2n}$, $(\bar\psi\psi)^{4m}$, and $(\bar\psi\psi)^{2l+4}$.

With the truncation of the effective action specified, we are now set to derive and integrate the flow equations. We resolve the coupled set of resummed anomalous dimensions $\eta_\psi$, $\eta_A$, and $\eta_\phi$, together with that of the parameters
$\big\{h,\bar \lambda_{\pm},\bar\lambda_{\rm VA},\lambda_{1\ldots n},\lambda^{(\tau)}_{1\ldots m},\lambda^{(\rho\tau)}_{1\ldots l}\big\}$. Their explicit flow equations are given in \cref{app:RGflows}.

\section{Critical coupling for $\dSSB$}
\label{sec:critical}
As previously discussed, the fermionic correlations in the effective action are sourced by the running of the singly (marginally) relevant gauge coupling. A useful quantity to define in this context is the minimal strength of the gauge coupling $\acrit$ required to trigger $\dSSB$. Studying this critical coupling as a function of the theory's symmetries or external parameters such as temperature and chemical potential provides direct information about the off-shell interplay underlying $\dSSB$ and allows us to delimit the phase structure~\cite{Gies:2005as,Kusafuka:2011fd,Terao:2007jm,Goertz:2024dnz,Li:2025tvu,Li:2026ayh}. 

$\acrit$ depends mainly on the purely fermionic sector and therefore can be computed independently of the detailed running of the gauge coupling and its respective dynamics. In practice, we determine it by integrating the flows to $k\to0$ for a range of fixed values of $\alpha_g$ and identify $\acrit$ as the largest value for which the fermionic sector remains in the symmetric phase, see e.g.~\cite{Goertz:2024dnz,Li:2025tvu,Li:2026ayh,Braun:2011pp} for more details.

\begin{figure}[t!]
	\centering
	\includegraphics[width=.425\columnwidth]{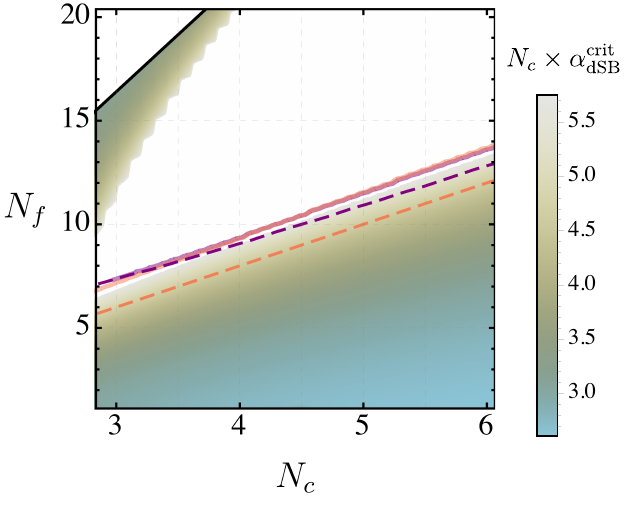}
	\hspace{0.1 cm}
	\raisebox{0.03\height}{\includegraphics[width=.525\columnwidth]{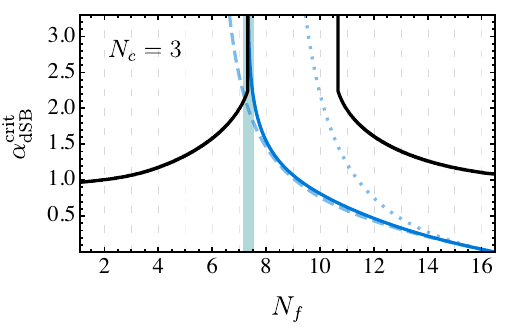}}
	\caption{In the left-hand plot, the critical gauge coupling $\acrit$ required to trigger $\dSSB$ is shown for different $N_c$ and $N_f$. Solid lines indicate theories for which $\acrit$ coincides with the fixed-point values $\alpha_g^*$ obtained from the three-loop (orange) and four-loop (purple) $\overline{{\rm MS}}$ beta functions. Dashed lines indicate theories for which $\alpha_g^*$ diverges.\\[0.25ex]
	In the right-hand plot, we project onto $N_c=3$ and show $\acrit$ as a black line, together with the two-loop (dotted), three-loop (dashed), and four-loop (solid) fixed-point values in blue for . The green shaded region denotes the critical region where $\eta_\psi\geq1$ evaluated at $\acrit$ and $\aFP$, see the discussion in \cref{sec:critical}.}
	\label{fig:alphacrit}
\end{figure}

In \cref{fig:alphacrit}, we show $\acrit$ for different $N_c$ and $N_f$ and project onto  $N_c=3$ in the right-hand panel. The critical coupling exhibits an important $N_f$ dependence and increases steadily up to the point where very strong dynamics are required for $\dSSB$  to occur.  This defines
\begin{align}
	\left.\Nfcrit\right|_{N_c=3}\simeq 7.30\,,
	\label{eq:Nfcrit}
\end{align}
as the critical number of flavours for a transition away from the few flavour phase with $\dSSB$. Importantly, \eqref{eq:Nfcrit} is self-consistently determined within the fRG and predicts the existence of a many flavour phase absent of $\dSSB$. 

For illustrative purposes, we can confront $\acrit$ with the CBZ fixed points obtained from multi-loop $\overline{{\rm MS}}$ beta functions~\cite{Herzog:2017ohr,Chetyrkin:2017bjc,vanRitbergen:1998pn,Ruijl:2017eht} and estimate the region where the conformal window can be found. The different $\aFP$ are shown on the right hand pannel of \cref{fig:alphacrit} by blue lines. When $\alpha_g^*<\acrit$, the fixed point is reached before the gauge interaction becomes strong enough to trigger $\dSSB$, and the theory remains conformal. Conversely, when $\alpha_g^*>\acrit$, $\dSSB$ occurs before the fixed point can be reached. Peculiarly, disappearance of $\acrit$ and the four-loop $\overline{\rm MS}$  fixed point into very strong values remarkable coincides at $\Nfcrit$.

Nonetheless, it is important to emphasize that any comparison of $\acrit$ and $\aFP$ should be regarded as an estimate and interpreted with some caution.  Perturbative results consistently capture the $N_c$ and $N_f$ dependence at each loop order and provide a powerful benchmark for the magnitude of $\aFP$. However, beyond two loops, the beta functions of marginal couplings become scheme-dependent, and so do the fixed-point values~\cite{Gracey:2023unc}. For example, using fixed points in the mMOM scheme~\cite{Gracey:2013sca,Gracey:2015uaa,Gracey:2022xjs}, we would obtain $\Nfcrit\simeq 6.8$ at both three and four loops. 

From the fRG side, the resummed structure of the flow equation incorporates contributions from arbitrarily high orders in the couplings. However, a consistent treatment of all $N_c$ and $N_f$ factors at high loop order would require all relevant loop topologies, including the momentum dependence of higher-order operators in the pure-gauge, gauge--fermion, and fermionic sectors. This would require a substantially more advanced truncation.

Furthermore, it is also convenient to contextualise with recent results from other non-perturbative approaches. Lattice simulations have progressively explored theories with different numbers of flavours~\cite{Hasenfratz:2023wbr,Hasenfratz:2022yws,Hasenfratz:2017qyr,Hasenfratz:2016dou,Hasenfratz:2020ess,LSD:2018inr,Appelquist:2011dp,Iwasaki:2003de,LatticeStrongDynamics:2018hun}, with evidence that $N_f=8$ lies just within the conformal window~\cite{Hasenfratz:2022zsa,DElia:2026oah,LatKMI:2016xxi,Appelquist:2007hu}. Recent Dyson--Schwinger equation studies~\cite{Chen:2026pqz}, which include important tensor structures in the fermion--gauge vertex, obtain a critical flavour number of $\simeq 6.8$. At the present stage, these two independent approaches are compatible with the results presented here.

Having established our main result, we proceed to examine its origin in \cref{sec:convergence,sec:propertiesdSB} and discuss its implications for the near-conformal regime in \cref{sec:conformalPT}.

\subsection{Non-perturbative corrections and convergence}
\label{sec:convergence}
\begin{figure}[t!]
	\centering
	\includegraphics[width=.3\columnwidth]{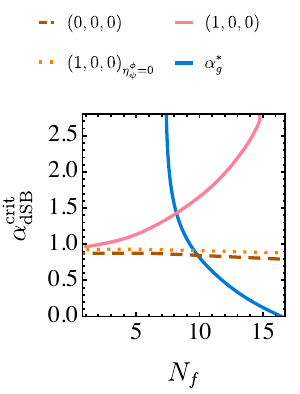}\hspace{-1.25cm}
	\includegraphics[width=.3\columnwidth]{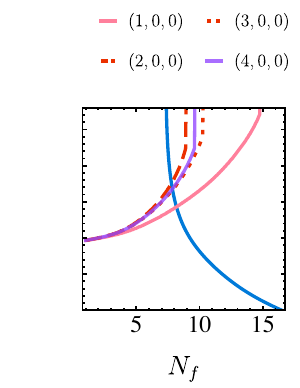}\hspace{-1.25cm}
	\includegraphics[width=.3\columnwidth]{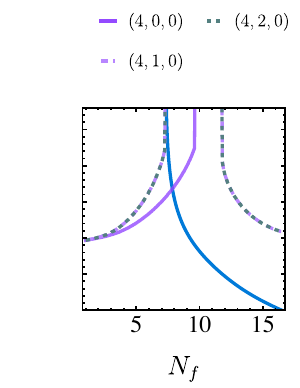}\hspace{-1.25cm}
	\includegraphics[width=.3\columnwidth]{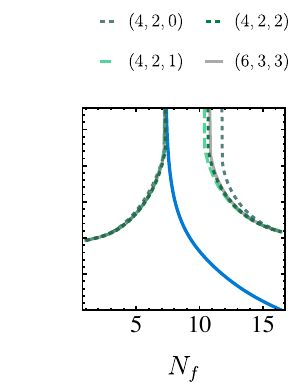}
	\caption{Convergence of $\acrit$ for different approximations and truncation orders, see \cref{sec:convergence}. The blue line shows the four-loop CBZ fixed point for reference. Each panel displays a different maximum order of the polynomial expansion \eqref{eq:potential}, labelled by $\left(N_{\rm max},\,M_{\rm max},\,L_{\rm max}\right)$.}
	\label{fig:alphacritconvergence}
\end{figure}

Within our setup, the convergence of $\acrit$ can be assessed by systematically improving the truncation of the effective action. This allows us to identify where the relevant off-shell dynamics responsible for the observed flavour dependence reside. In \cref{fig:alphacritconvergence}, we show the resulting $N_f$ dependence of $\acrit$ for a sequence of truncations, isolating the effects of bosonisation, higher-order interactions, and additional invariants.

\paragraph{\textit{Momentum dependencies via bosonisation.}}
As discussed in \cref{sec:dynamical bosonisation}, dynamical bosonisation conveniently gives access to both higher-dimensional fermionic interactions through the effective potential and the momentum dependence of the selected four-fermion channel. This goes beyond the point-like approximation and is particularly important here because it captures loop topologies that carry the leading $N_f$ corrections. These arise through the flow of the Yukawa coupling in the fermion anomalous dimension,
\begin{align}\label{eq:etapsiexplicit}
	\eta_\psi= -\frac{\partial_t Z_\psi}{Z_\psi}= \frac{g^2}{16 \pi^2}\frac{(N_c^2-1)}{2 N_c }\left(\frac{\eta_\psi}{4}-\frac{\eta_A}{5}\right)+\frac{h^2 N_f }{16 \pi^2}\left(1-\frac{\eta_\phi}{5}\right)\,,
\end{align}
as derived in \eqref{eq:etapsi}. The first term arises from gauge-field-mediated diagrams and is proportional only to anomalous dimensions, which themselves scale with powers of $\alpha_g$. The second comes from the meson exchange and scales as $h^2\propto\alpha_g^2$, since the gauge-mediated box diagrams generate the analogous four-fermion interactions. Thus, the corrections here enter beyond one loop as expected in Landau gauge.

The bosonic contribution captures the effect of the dominant scalar-pseudoscalar channel, whose modes become critical as $\dSSB$ sets in. Although this correction is formally two-loop and coupling suppressed, it is leading in $N_f$ due to of the multiplicity of the resonant modes and becomes dominant in the many flavour limit. Consequently, this term is essential for transmitting the off-shell dynamics associated with $\dSSB$ into $\eta_\psi$ and the effective action.

It is worth stressing that $\eta_\psi$ is obtained from a derivative expansion of the fermion two-point function in the external momentum and evaluated at vanishing momentum, see \cref{app:anomdim}. In the point-like four-fermion formulation, the coupling is momentum-independent by construction, and this contribution therefore vanishes identically. Bosonisation removes this limitation by representing the momentum dependence of the selected four-fermion channel through the bosonic dispersion. The corresponding Yukawa loops then become non-vanishing and generate the relevant $N_f$ correction.

The impact of these corrections on $\acrit$ is shown in the leftmost panel of \cref{fig:alphacritconvergence} where we compare the point-like approximation (brown dashed line), first studied in \cite{Gies:2005as}, with the lowest-order bosonised truncation ($N_{\rm max}=1$), both with (red solid line) and without (orange dotted line) the bosonic contribution to $\eta_\psi$. The comparison shows that the momentum dependence generated by bosonisation is essential for recovering the correct leading $N_f$ dependence.

One may then ask whether additional leading corrections enter $\eta_\psi$ through the tensor structures that remain unbosonised. Of these, only the (V+A) channel contributes to the scalar trace of the fermion two-point function, with a contribution proportional to $-\bar{\lambda}_{+}/(N_cN_f)$. This channel is subleading and suppressed in both colour and flavour, so no additional leading $N_f$ corrections are expected. We have also checked this explicitly by including its contribution in the anomalous-dimension flow. As it is discussed below, these subdominant tensor structures do not affect $\acrit$ for $N_f<\Nfcrit$.

\paragraph{\textit{Multi-fermion interactions and flavour symmetry.}}
Further leading $N_f$ dependencies arise from multi-fermion interactions encoded in the mesonic effective potential. These contributions can be analysed from the flow of the small-field curvature around the symmetric vacuum, given by the bosonic curvature mass and represented by the lowest-order coefficient $\lambda_1$ in the $\bar\rho$ expansion,
\begin{align}	\label{eq:flowlambda1}
	\partial_t (\partial_{\bar\rho} u(\bar\rho,\,\bar{\tau}))_{\bar\rho=\bar\tau=0}\equiv \partial_t \lambda_1
	=&(-2+\eta_\phi) \lambda_1
	+\frac{h^2 N_c}{16 \pi ^2}\left(1-\frac{\eta_\psi}{5}\right)
	\\[0.5ex]
	&\hspace{2cm}
	-\frac{N_f^2-1}{32 \pi ^2 N_f (\lambda_1+1)^2}
	\left(\lambda_2 N_f +2 \lambda^{(\tau)}_1\right)
	\left(1+\frac{\eta_\phi}{6}\right)\,.
	\notag
\end{align}
This flow is particularly important because the transition to the broken phase is signalled by the small-field curvature becoming negative, $\lambda_1<0$, at finite $k$. The first two terms describe the canonical and anomalous scaling of the curvature and the fermionic contribution to the bosonic two-point function, respectively. The last term arises from mesonic self-interactions in the two invariants included.

Including higher-order interactions in $\bar\rho$, the $\acrit$ profile converges rapidly, as shown in the second panel of \cref{fig:alphacritconvergence}. Interactions involving the $\bar\tau$ invariant first enter through the bosonic four-point functions. Including these interactions introduces a further enhancement of the $N_f$ dependence, as shown in the third panel. As emphasised above, $\bar\tau$ is essential for reproducing the correct symmetry-breaking pattern and without it, the fermionic sector acquires an enlarged $SO(2N_f)$ symmetry instead of the physical $U(N_f)_L\times U(N_f)_R$. The difference between the converged results in the second and third panels should therefore be understood primarily as a consequence of realising the correct flavour symmetry. Finally, interactions combining the two invariants have only a mild impact, as shown in the right-most panel.

In conclusion, we find that the $N_f$ dependence of $\acrit$ is well converged for a polynomial potential truncated at order $(4,2,2)$, even in the non-perturbative regime where strong gauge couplings are required to trigger $\dSSB$. Up to this order, all interactions through $(\bar\psi\psi)^8$ in the bosonised channel are included. Beyond this order, additional invariants, such as ${\rm inv}_3$ combined with $\bar\rho$, begin to contribute, but their higher canonical dimension makes them strongly suppressed.

\subsection{$N_f$ scaling of the effective action}
\label{sec:propertiesdSB}
\begin{figure}[t!]
	\centering
	\includegraphics[width=\columnwidth]{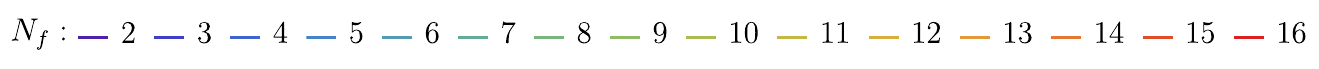}
	\includegraphics[width=.32\columnwidth]{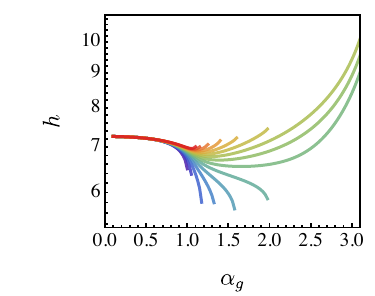}
	\includegraphics[width=.32\columnwidth]{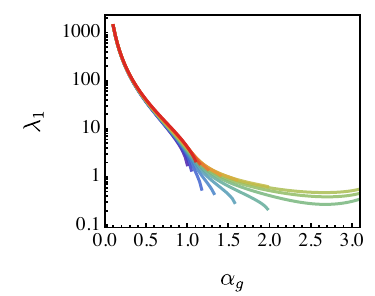}
	\includegraphics[width=.32\columnwidth]{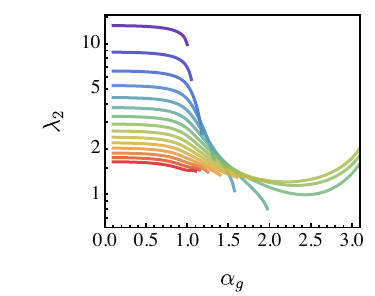}
	\includegraphics[width=.32\columnwidth]{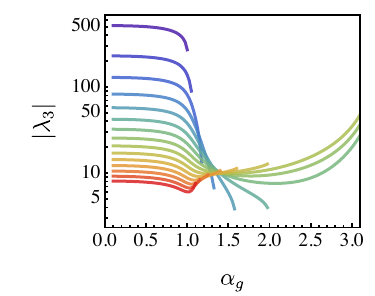}
	\includegraphics[width=.32\columnwidth]{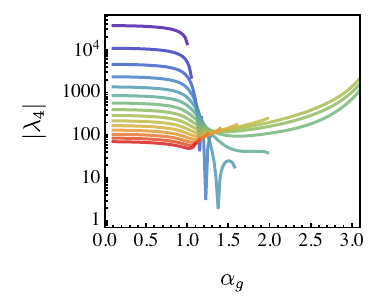}
	\includegraphics[width=.32\columnwidth]{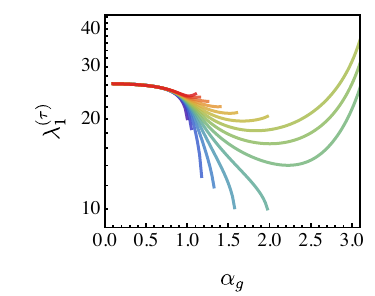}
	\includegraphics[width=.32\columnwidth]{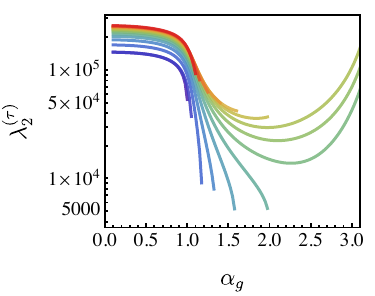}
	\includegraphics[width=.32\columnwidth]{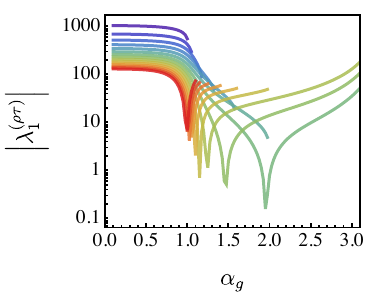}
	\includegraphics[width=.32\columnwidth]{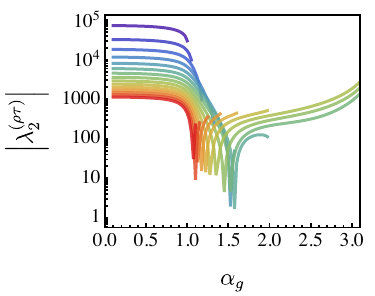}
	\includegraphics[width=.32\columnwidth]{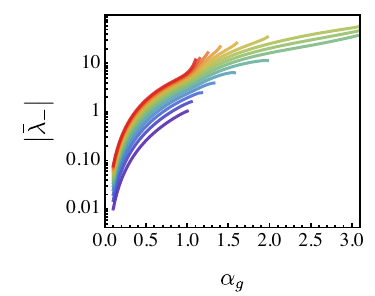}
	\includegraphics[width=.32\columnwidth]{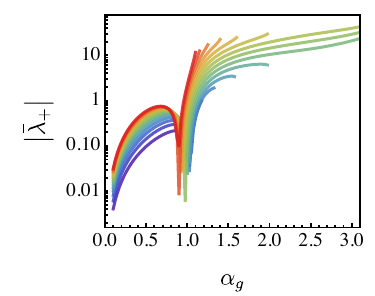}
	\includegraphics[width=.32\columnwidth]{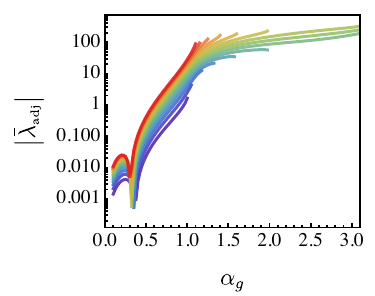}
	\caption{Couplings of the effective action, obtained by integrating the RG flows, as a function of $\alpha_g$ for different $N_f$.}
	\label{fig:couplingsdiffalphagNf}
\end{figure}

Having established which are the non-perturbative corrections that shape the $\acrit$ profile, we now examine the resulting fermionic effective action in more detail. In \cref{fig:couplingsdiffalphagNf}, we show the couplings obtained after integrating the RG flows for different $N_f$ and fixed values of $\alpha_g$. All couplings remain well behaved as $N_f$ is increased, including in the region above $\Nfcrit$ where very strong gauge dynamics are reached. In particular, the parameters of the effective potential remain stable even at strong coupling. A notable feature is that, for many flavours and large $\alpha_g$, the potential becomes increasingly flat. Although some expansion coefficients change sign, this is a feature of the polynomial parametrisation rather than an indication of a new minimum. No additional minimum develops, as shown explicitly for $N_f=8$ in \cref{fig:effectiveVNf8}.

\begin{figure}[t!]
	\centering
	\includegraphics[width=.275\columnwidth]{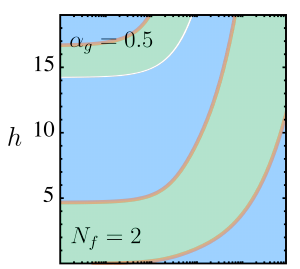}\hspace{-.75cm}
	\includegraphics[width=.275\columnwidth]{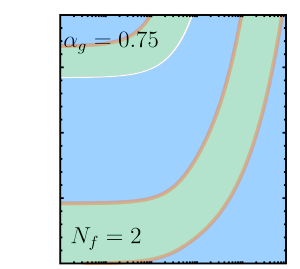}\hspace{-.75cm}
	\includegraphics[width=.275\columnwidth]{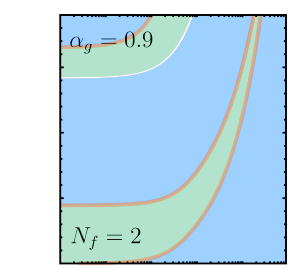}\hspace{-.75cm}
	\includegraphics[width=.275\columnwidth]{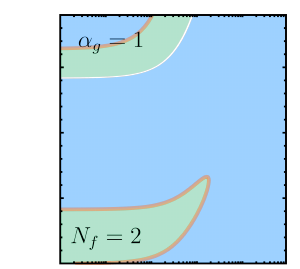}\hspace{-.75cm}
	
	\includegraphics[width=.275\columnwidth]{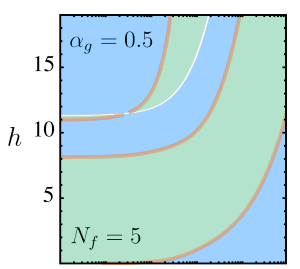}\hspace{-.75cm}
	\includegraphics[width=.275\columnwidth]{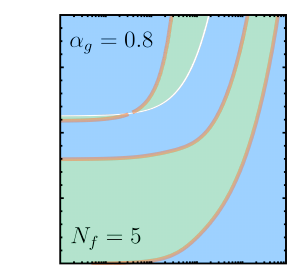}\hspace{-.75cm}
	\includegraphics[width=.275\columnwidth]{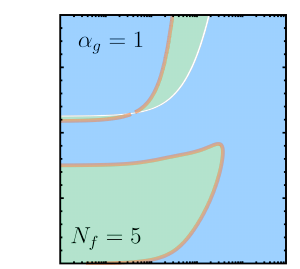}\hspace{-.75cm}
	\includegraphics[width=.275\columnwidth]{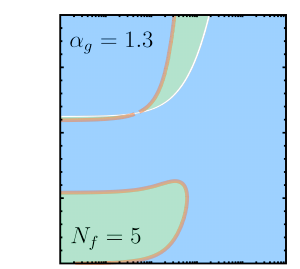}\hspace{-.75cm}
	
	\includegraphics[width=.275\columnwidth]{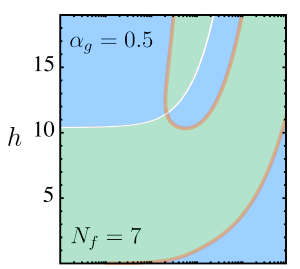}\hspace{-.75cm}
	\includegraphics[width=.275\columnwidth]{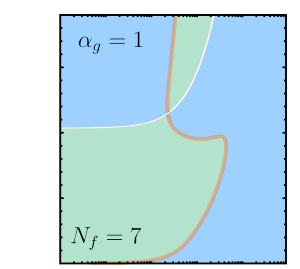}\hspace{-.75cm}
	\includegraphics[width=.275\columnwidth]{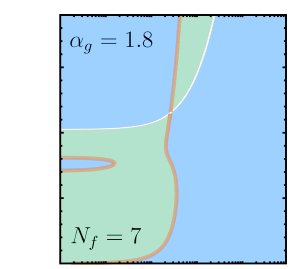}\hspace{-.75cm}
	\includegraphics[width=.275\columnwidth]{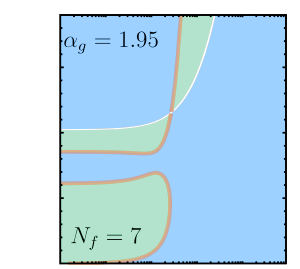}\hspace{-.75cm}
	
	\includegraphics[width=.275\columnwidth]{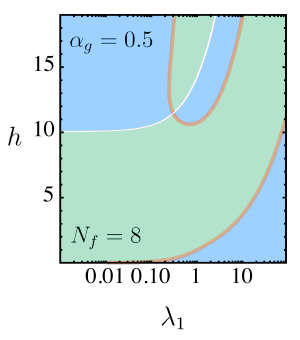}\hspace{-.75cm}
	\includegraphics[width=.275\columnwidth]{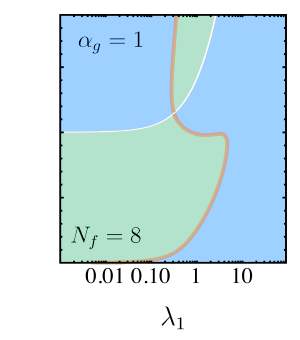}\hspace{-.75cm}
	\includegraphics[width=.275\columnwidth]{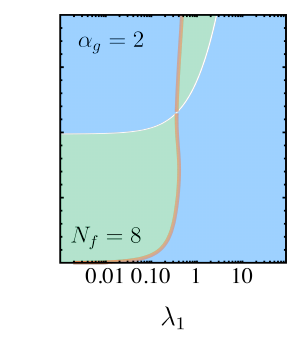}\hspace{-.75cm}
	\includegraphics[width=.275\columnwidth]{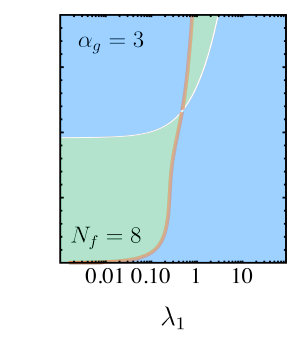}\hspace{-.75cm}
	
	\caption{The coloured surface shows the sign of the flow of the composite coupling $\partial_t(h^2/\lambda_1)$ as a function of the bosonised couplings, evaluated along the integrated flows. Green indicates a positive flow and blue a negative flow. The orange line denotes a root of the flow, while the white line marks the location of a strong-coupling pole.}
	\label{fig:FPmerger}
\end{figure}

To further understand the transition between symmetric and broken phases, we analyse the flow of $h$ and $\lambda_1$ through the composite coupling $\epsilon_\phi=h^2/\lambda_1$~\cite{Gies:2001nw,Gies:2002hq}, which can be viewed as an improved four-fermion coupling similar to \eqref{eq:lambdaEoMphi}. In \cref{fig:FPmerger}, we display the structure of this flow, showing its roots (orange lines), pole (white line), and regions of positive (green) and negative (blue) flow.

In the weak-coupling regime, $\alpha_g\ll1$, the composite coupling starts in the bottom-right corner, $\epsilon_\phi\ll1$, and flows towards larger values, approaching the weakest root of the flow. In the top row, we show the case for $N_f=2$, where the standard fixed-point merger described in \cref{sec:effectiveaction} applies. As the gauge dynamics strengthen, the theory flows towards smaller values of $\lambda_1$, corresponding to a flattening of the mesonic potential. For $\alpha_g\lesssim\acrit$, the two fixed-point branches merge, after which the flow continues into the broken phase, with $\lambda_1\to0$ and $h\sim5$.

As $N_f$ increases, a fixed point initially located at large $\epsilon_\phi$ moves towards smaller $h$ and merges with one of the two root branches, even in the weak-gauge-coupling regime. For $N_f\simeq7$, stronger gauge dynamics can still undo this merger and trigger $\dSSB$, as shown in the third row. For $N_f\gtrsim\Nfcrit$, however, the merger can no longer be undone, as shown in the fourth row. Instead, the theory flows towards a stable point where the root and pole of the flow coexist. This prevents $\dSSB$ and leads to the sharp increase of $\acrit$, while the flow itself remains moderate and the couplings of the effective action stay under control, as shown in \cref{fig:couplingsdiffalphagNf}.

The pole in the flow of $\epsilon_\phi$ originates in $\eta_A$ and enters through the resummed $\eta_\psi$ contribution to the flow of $h$. This singularity is associated with the onset of colour confinement and can be resolved by accounting for the dynamical generation of a gauge mass gap, which is not included in the present truncation. Theories with $N_f\gtrsim\Nfcrit$ should therefore be interpreted as lying in a regime where stability against $\dSSB$ persists while confining dynamics set in.

\paragraph{\textit{Other $\dSSB$ patterns?}} 
In the last row of \cref{fig:couplingsdiffalphagNf}, we show the four-fermion couplings of the subdominant tensor structures. While these grow as $\alpha_g$ increases, they do not affect the determination of $\acrit$ for $N_f<\Nfcrit$ at all. Nonetheless, for $N_f\gtrsim11$, in the regime where the scalar-pseudoscalar channel does not allow for $\dSSB$, we find a divergence in $\bar\lambda_{\rm adj}$ occurs, indicating the formation of a condensate in a different channel from the standard one and hence a new symmetry-breaking pattern.

This rather exotic fixed-point merger in the (V--A)$_{\rm adj}$ channel results in the  $\acrit$ many-flavour branch appreciable in \cref{fig:alphacrit}. This lies well within the region where the CBZ fixed point is reliably found and therefore will not be reached by the gauge dynamics. Moreover, this result should be interpreted with caution, since the non-(S--P) channels have not been bosonised and their flows therefore include only the lowest-order point-like corrections which usually suffice to reproduce the qualitative picture.

\section{Phase structure and quantum phase transition}
\label{sec:conformalPT}
We have established that $\acrit$ develops a strong leading dependence on $N_f$, implying that increasingly strong gauge interactions are required to trigger $\dSSB$ as the number of flavours increases. This behaviour leads to a qualitatively different picture of the near-critical region and of the quantum phase transition into the conformal window.

\subsection{Absence of walking and order of the quantum phase transition}\label{sec:absencewalking}
The near-conformal scaling of observables can be inferred from the relative behaviour of $\acrit$ and $\aFP$. If theories just below the lower boundary of the conformal window satisfy $\acrit\lesssim\aFP$, the gauge coupling remains close to the IR fixed point over an extended range of scales before $\acrit$ is reached and $\dSSB$ occurs. This leads to the appearence of \textit{walking regimes} as a product of the CBZ fixed point screening the confining dynamics while remaining sufficiently strong to trigger $\dSSB$. In this regime, the gauge CBZ fixed point coexists with a near-critical fixed-point merger in the four-fermion couplings.

Since in walking theories the highest dynamical scale is set by that at which the four-fermion fixed-point merger occurs, its disapperence towards the conformal window can be shown to be continuous and of BKT-type~\cite{Berezinskii:1970pzv,Kosterlitz:1973xp}. In this region, the $\dSSB$ scale ($\kSSB$) and related observables scale exponentially as the transition at $\Nfcrit$ is approached from lower flavour numbers. This behaviour is commonly referred to as Miransky scaling~\cite{Miransky:1989qc} and has been computed in several works~\cite{Braun:2010qs,Braun:2009ns,Goertz:2024dnz}.

\begin{figure}[t!]
	\centering
	\includegraphics[width=.8\columnwidth]{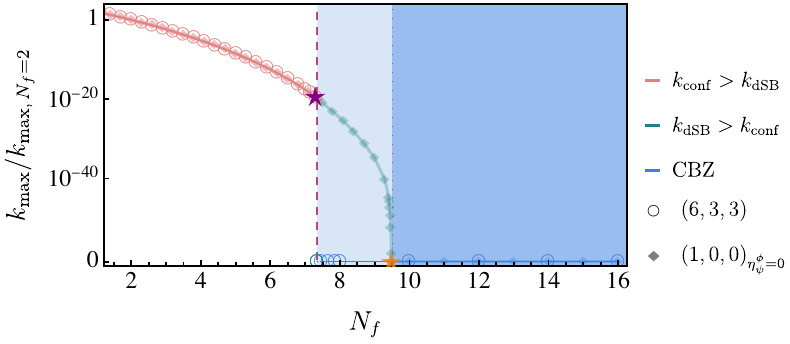}
	\caption{First dynamically generated scale for $N_c=3$ as a function of $N_f$, normalised to the two-flavour case. We compare two approximations: the present non-perturbative computation (open circles) and the result obtained using only the lowest-order point-like four-fermion interactions (filled diamonds). Three regimes are indicated: one in which the highest dynamically generated scale is set by colour confinement (red), one in which $\dSSB$ occurs and leads to walking (green), and the conformal regime (blue).}
	\label{fig:conformalPT}
\end{figure}

The present results instead point towards a more subtle scenario, in which both $\acrit$ and $\aFP$ increase and become large near the same $\Nfcrit$. In this case, there is no weakly coupled regime in which $\acrit\lesssim\aFP$ and the gauge dynamics responsible for confinement cannot be screened. Consequently, walking regimes and Miransky/BKT scaling cannot be established in the near-conformal region. Instead, this scenario points towards a first-order quantum phase transition between the dynamical and conformal phases.

To illustrate this distinction, we integrate the flow equations together with the running gauge coupling obtained from the four-loop $\overline{\mathrm{MS}}$ $\beta$-function. This allows us to estimate the confinement scale ($\kconf$) by tracking the IR singularity of the gauge coupling, as well as the onset of the CBZ window. In \cref{fig:conformalPT}, we show the highest dynamically generated scale for $N_c=3$ as a function of $N_f$, comparing the two approximations discussed above. 

Three regimes can be distinguished. For few flavours, the first divergence occurs in the pure gauge sector, indicating that confinement sets the highest dynamical scale (red markers). In the point-like four-fermion approximation, $\Nfcrit$ is instead larger and $\acrit\lesssim\aFP$ is realised over an intermediate range. In this regime, $\dSSB$ sets the highest dynamical scale, giving rise to Miransky scaling (green diamonds). Finally, for sufficiently many flavours (blue markers and shaded region), $\aFP\ll1$, theories lie well inside the CBZ window and no scale is dynamically generated.

While the two approximations agree well in the two asymptotic regimes, they differ qualitatively in the near-conformal region. The present calculation points towards a first-order quantum phase transition (red star), whereas the lower-order point-like approximation supports a continuous, Miransky-like transition (orange star).

We emphasise that, although our best approximation, combining the four-loop fixed-point profile with the converged determination of $\acrit$, supports the absence of walking, this cannot be excluded. For this analysis the CBZ fixed point is required and three-loop results or alternative renormalisation schemes could allow a region in $N_f$ where $\acrit<\alpha_g^*$. However, the observed growth of $\acrit$ suggests that, if such a regime exists, it is likely to occur at strong coupling and therefore confining dynamics are not efficiently screened. A conclusive determination again requires a self-consistent calculation of both the fixed-point profile and $\acrit$ within the same framework.

\subsection{The critical region and a symmetric mass gap}\label{sec:criticalregion}
The large value of $\acrit$ in the critical region raises the question of whether other non-perturbative dynamics could enter the game. While our analysis shows that $\dSSB$ becomes increasingly difficult to trigger colour confinement is still be expected to occur.

Confining dynamics have a finite maximal strength set by the dynamical generation of a gauge mass gap. For comparison purposes, the Landau gauge quark--gluon exchange coupling in two-flavour QCD peakes at values of order $\alpha_g\sim2.5$~\cite{Ihssen:2024miv,Goertz:2024dnz,Gholami:2026cgd}. Na\"ively extrapolating and assuming no $N_f$ dependence of the critical coupling for confinement, one cannot exclude a phase with colour confinement but without $\dSSB$ near $\Nfcrit$.

This phenomenon has previously been discussed in chiral gauge theories~\cite{Eichten:1985fs,Karasik:2022gve,Li:2026ayh,Tong:2021phe}, where perturbative 't~Hooft anomalies can be matched by massless baryons in the IR~\cite{Bars:1981se}. Similar dynamics have also been suggested in QCD-like theories with many flavours~\cite{Chen:2026pqz,Witzel:2024bly,Hasenfratz:2023wbr,Hasenfratz:2022zsa}. However, in the present case the 't~Hooft anomaly matching conditions impose strong constraints on the possible IR spectrum~\cite{Ciambriello:2022wmh,Tanizaki:2018wtg,Hasenfratz:2026orf}. The existence of such a phase therefore remains speculative, as neither suitable massless degrees of freedom nor a dynamical mechanism for generating them is so far known.

Along these lines, our results indicate that the strong dynamics lead to peculiar features in the critical region. Consider the chiral fermion two-point function,
\begin{align}
	\Gamma_k^{(\bar\psi\psi)}(p)
	&= Z_{\psi,k}(p)\,{\rm i}\slashed{p},
	&&\text{satisfying}&&
	\lim_{p\to0}
	\big|\Gamma_{k\to0}^{(\bar\psi\psi)}(p)\big| ^{-1}<\infty \,,
	\notag
\end{align}
such that the propagator is gapped without the appearance of a symmetry-breaking mass term and which allows additional Lorentz-invariant structures to appear. A sufficient condition for such a symmetric gap is therefore
\begin{align}
	Z_{\psi,k\to0}(p)\propto p^{-(1+\varepsilon)}
	&&{\rm with}\quad
	\varepsilon\geq0,&&
	&&{\rm and }&&
	\eta_{\psi,k\to0}(p=0)\geq1.
	\notag
\end{align}
Such behaviour is rarely realised in fermionic systems because it requires very strong dynamics, which typically trigger $\dSSB$ first. In the present setting, however, the onset of $\dSSB$ is blocked in the large-$N_f$ region and we find
\begin{align}
	\left.\eta_{\psi}\right|_{\acrit}\geq1
	&&\text{for}&&
	N_f>7.2,
	&&\text{and}&&
	\left.\eta_{\psi}\right|_{\aFP}\geq1
	&&\text{for}&&
	N_f<
	\left(
	\left.7.3\right|_{3{\rm\tiny loop}},\,	\left.7.6\right|_{4{\rm\tiny loop}}
	\right),
	\notag
\end{align}
which is precisely the region associated with the transition to the conformal window highlighted in green in \cref{fig:alphacrit} (see also the anomalous dimensions shown in \cref{fig:etas}). This suggests that the critical region may support a non-perturbative fermionic symmetric mass gap in a colour-confining phase. However, as discussed above, this possibility is constrained by the need for a consistent IR spectrum which satisfies the UV anomaly constraints which is not straightforwardly fullfilled here. A detailed investigation of this possibility and its properties along the inclusion of confining dynamics along the lines of~\cite{Goertz:2024dnz} is left for future work.

As a final comment, our results indicate that the condition $\acrit=\aFP$ occurs at strong coupling. This implies a lower $\Nfcrit$ and the existence of theories with strongly coupled fixed points within the CBZ window. Such theories provide interesting test cases which study within the present framework, for instance determining the conformal scaling dimensions.

\section{Conclusions}
\label{sec:conclusions}
In this work, we have studied the emergence of dynamical symmetry breaking ($\dSSB$) in QCD-like theories with many fermion flavours $N_f$. We employed the effective action formalism and the functional Renormalisation Group (fRG) to include non-perturbative corrections, particularly in the fermionic sector where the relevant off-shell dynamics reside.

For this computation we used the generalised flow equation to implement scale-dependent field redefinitions and treat high-order fermionic self-interactions in the dominant scalar-pseudoscalar channel in the form of dynamical bosons. This upgrade allows us to account for momentum dependencies beyond the point-like approximation and to include higher-order fermionic self-interactions. We also included the subleading mesonic field invariant that is essential to implement correctly the chiral symmetry and its breaking pattern.

We determined the critical strength of the gauge dynamics, $\acrit$, required to trigger $\dSSB$. The newly included corrections introduce a strong $N_f$ dependence and lead to a qualitatively new picture of $\dSSB$ across the phase diagram. In particular, we derive the critical number of flavours $N_f^{\rm crit}\sim7.30$ for $N_c=3$, beyond which $\acrit$ increases sharply indicating the existence of a phase without $\dSSB$. Remarkably, this feature is obtained purely and self-consistently from the flow computation and does not rely on an estimate of the CBZ fixed point. It nevertheless coincidentally agrees with the number of flavours at which the four-loop $\overline{{\rm MS}}$ fixed point disappears to infinity.

Within this setup, we systematically analysed the impact of the included quantum corrections, namely higher-order interactions, additional invariants, and momentum dependencies, and studied the effective action as $N_f$ and $\alpha_g$ are increased. By analysing the RG flow of the bosonised sector, we find that the mechanism underlying $\dSSB$ changes from a fixed-point merger at small $N_f$ to a regime in which the transition to the broken phase is prevented by the appearence of a stable root.

The new $N_f$ dependence of $\acrit$ has important implications for the near-critical regime and the quantum phase transition to the conformal window. In particular, its strong flavour dependence points towards the absence of Miransky or BKT scaling and consequently walking regimes. Instead, our results favour a first-order quantum phase transition with a critical region in which novel dynamics, including confinement without $\dSSB$ and symmetric fermion mass gaps, may occur. While we find strong indications for this, establishing it requires a treatment incorporating confining dynamics and reproducing an IR spectrum which is compatible with the UV anomalies. We will explore this peculiar regime in a future work.

\section*{Acknowledgements}
I would like to thank Holger~Gies, John~A.~Gracey, Joerg~Jaeckel and Cenke~Xu for discussions, and to Jan~M.~Pawlowski, Fabian~Rennecke, Franz~R.~Sattler and Shahram~Vatani also for collaborations on related topics.  APG is supported by the RIKEN Special Postdoctoral Researcher (SPDR) Program and the RIKEN Incentive Research Grant.

\section*{Appendices}
\appendix

\section{Definition of the effective action}
\label{app:Effective action}
In this appendix, we provide further details on the effective action, with particular emphasis on the fermionic sector, which is central to the present work.

The pure gauge sector drives the dynamics. Rather than solving its flow independently, we treat the gauge coupling as an external source for the fermionic correlations. The gauge sector enters our analysis in two ways. First, we estimate the onset of the CBZ fixed point using high-order perturbative $\beta$-functions. Second, the flows in the fermionic sector, as well as the meson and fermion anomalous dimensions, depend on the gauge-boson anomalous dimension. The latter is computed within the truncation described in the appendices of~\cite{Goertz:2024dnz}, with its derivation and explicit form given in \cref{app:anomdim}.

\paragraph{Gauge--fermion effective action.}
An important component of the effective action is the gauge--fermion sector, which connects the pure gauge and fermionic correlations,
\begin{align}
	\Gamma_{{\rm gauge-fermion},\,k}\left[A_\mu,\bar\psi, \psi\right]
	= \int_p Z_{\psi}\,\bar\psi\,\gamma_\mu
	\left(
	\partial_\mu
	+ g\,Z_A^{1/2}\,T_c^a A_\mu^a
	\right)\psi\,.
	\label{eq:DiracAction}
\end{align}
Here we retained only the leading tensor structure and momentum dependence of the vertex. The generators $T_c^a$ are taken in the fundamental representation of the colour gauge group as in physical QCD.

\paragraph{Four-fermion interactions and tensor structures.}
The Fierz-complete basis of momentum-independent four-fermion operators compatible with $U(N_f)_L\times U(N_f)_R$ and $SU(N_c)$ gauge invariance can be arranged into four independent tensor structures,
\begin{subequations}
	\label{eq:4FermiTensors}
	\begin{align}
		{\cal T}_{({\rm V-A})}
		&=
		\left(\bar\psi\gamma_\mu T_f^0\psi\right)^2
		+\left(\bar\psi\gamma_\mu\gamma_5 T_f^0\psi\right)^2\,,
		\\[1ex]
		{\cal T}_{({\rm V+A})}
		&=
		\left(\bar\psi\gamma_\mu T_f^0\psi\right)^2
		-\left(\bar\psi\gamma_\mu\gamma_5 T_f^0\psi\right)^2\,,
		\\[1ex]
		{\cal T}_{({\rm V-A})}^{\rm adj}
		&=
		\left(\bar\psi\gamma_\mu T_f^0 T_c^a\psi\right)^2
		+\left(\bar\psi\gamma_\mu\gamma_5 T_f^0 T_c^a\psi\right)^2\,,
	\end{align}
	and the dominant scalar-pseudoscalar tensor structure
	\begin{align}
		{\cal T}_{({\rm S-P})}
		={\cal T}_\sigma+{\cal T}_\pi+{\cal T}_a+{\cal T}_\eta\,,
	\end{align}
	with
	\begin{align}
		&{\cal T}_\sigma
		=\left(\bar\psi\,\mathbbm{T}_\sigma\psi\right)^2,
		&&
		{\cal T}_\pi
		=\left(\bar\psi\,\mathbbm{T}_\pi\psi\right)^2,
		&&
		{\cal T}_a
		=\left(\bar\psi\,\mathbbm{T}_a\psi\right)^2,
		&&
		{\cal T}_\eta
		=\left(\bar\psi\,\mathbbm{T}_\eta\psi\right)^2,
		\label{eq:AxialSplitTSP}
	\end{align}
	where
	\begin{align}
		&\mathbbm{T}_\sigma=T_f^0,
		&&
		\mathbbm{T}_\pi={\rm i}\gamma_5T_f^a,
		&&
		\mathbbm{T}_a=T_f^a,
		&&
		\mathbbm{T}_\eta={\rm i}\gamma_5T_f^0.
		\label{eq:mathbbmT}
	\end{align}
\end{subequations}
We note that this basis is Fierz-complete only for $N_c\geq3$. For $N_c=2$, the pseudo-reality of the fundamental representation gives rise to Pauli--G\"{u}rsey symmetry, which relates quarks and diquarks. An additional Fierz identity then renders ${\cal T}_{({\rm V-A})}^{\rm adj}$ linearly dependent on the remaining three tensor structures and the complete basis instead includes a diquark channel \cite{Fejos:2026tyd,Fejos:2025nvd,Gholami:2026cgd,Khan:2015puu,Braun:2009gm,Berges:1998rc}. For completeness, we have also computed the critical coupling for $N_c=2$ after removing the redundant tensor structure but without including the diquark one.  Here we find
\begin{align}
	\left.N_f^{\rm crit}\right|_{N_c=2}\simeq 5.11\,,
\end{align}
purely determined by the divergence of $\acrit$.

\section{Flow equations}
\label{app:RGflows}
All flow equations are derived from \labelcref{eq:GenfRG} by taking the appropriate functional derivatives of the effective average action and projecting onto the respective tensor structures. We evaluate them at vanishing external momenta in a symmetric momentum configuration and in the Landau gauge. The symbolic forms of the diagrammatic flow equations were derived using existing automated software packages~\cite{Huber:2019dkb,Pawlowski:2021tkk,Sattler:2026csm}.

Furthermore, we employ the four-dimensional Litim regulator \cite{Litim:2001up},
\begin{align}
	R_k^\phi(p^2)
	&=Z_\phi(k^2-p^2)\theta(k^2-p^2),
	&
	R_k^\psi(p^2)
	&=Z_\psi\,\slashed p
	\left(\sqrt{k^2/p^2}-1\right)\theta(k^2-p^2),
	\label{eq:regulators}
\end{align}
for bosonic and fermionic fields, respectively.

To solve the system of RG flows, we need to specify boundary conditions for the parameters. Only the relevant deformations of the critical surface need to be fixed, namely the gauge coupling, which sets the scales in the theory. The remaining parameters correspond to canonically irrelevant operators and are predicted by the single marginally relevant parameter. In other words, these operators are generated and determined by $\alpha_g$. Their independence of the boundary conditions is well established and has been demonstrated in several setups, see e.g.~\cite{Goertz:2024dnz,Gholami:2026cgd}.

\subsection{Anomalous dimensions}
\label{app:anomdim}
In the present setup we have three fields for which we derive their anomalous dimensions by projecting onto the respective dispersion terms in a derivative expansion in momentum. 

In the symmetric regime, the mesonic modes are degenerate, and the single anomalous dimension reads
\begin{align}
	\eta_\phi
	&=-\frac{\partial_t Z_\phi}{Z_\phi}
	=-
	\left.
	\frac{
		\partial_{p^2}
		\left(
		{\cal P}^{(\phi\phi)}
		\partial_t\Gamma_k^{(\phi\phi)}
		\right)
	}{
		Z_\phi\,{\rm Tr}
		\left[
		{\cal P}^{(\phi\phi)}
		{\cal P}^{(\phi\phi)}
		\right]
	}
	\right|_{p=0}
	=
	\left(1-\frac{\eta_\psi}{4}\right)
	\frac{h^2N_c^2}{8\pi^2}\,,
	\label{eq:etaphi}
\end{align}
where ${\cal P}^{(\phi\phi)}$ denotes the projection operator including the appropriate flavour and tensor structures for each bosonic mode. In the symmetric regime, the only contribution arises from the fermion polarisation diagram.

For the gauge field, the anomalous dimension reads
\begin{align}
	\eta_A
	&=-\frac{\partial_t Z_A}{Z_A}
	=-
	\left.
	\frac{
		\partial_{p^2}
		\left(
		\delta_c^{ab}\Pi^\perp_{\mu\nu}
		\partial_t\Gamma_k^{(AA)}
		\right)
	}{
		3(N_c^2-1)Z_A
	}
	\right|_{p=0}=
	-\left(156-31\eta_A\right)
	\frac{g^2N_c}{576\pi^2}
	+\left(1-\frac{\eta_\psi}{4}\right)
	\frac{g^2N_f}{12\pi^2}\,,
	\label{eq:etaA}
\end{align}
where we project onto the transverse mode of the two-point function and contract the colour structure in the adjoint representation.

Finally, the fermion anomalous dimension reads
\begin{align}
	\eta_\psi
	&=-\frac{\partial_t Z_\psi}{Z_\psi}
	=-
	\left.
	\frac{
		\partial_{p^2}
		\left(
		\delta_f^{ij}\delta_c^{ij}\,
		{\rm i}\slashed p\,
		\partial_t\Gamma_k^{(\bar\psi\psi)}
		\right)
	}{
		4N_cN_fZ_\psi
	}
	\right|_{p=0}\,,
	\label{eq:etapsi}
\end{align}
where the explicit expression is given in \eqref{eq:etapsiexplicit}.

Collecting these three expressions, we solve the system of anomalous dimensions and obtain their resummed form which includes higher-order effects. These are then employed in solving the complete system of flow equations.
\begin{figure}[t!]
	\centering
	\includegraphics[width=.3\columnwidth]{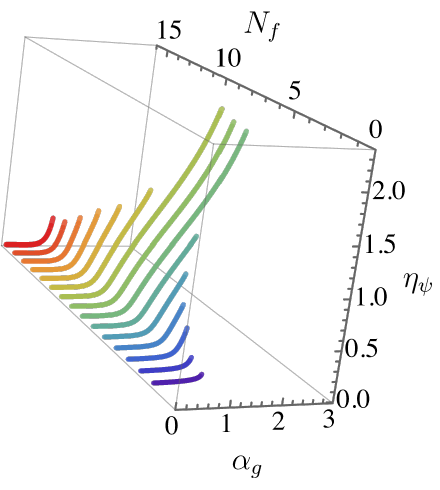}\hspace{1cm}
	\includegraphics[width=.27\columnwidth]{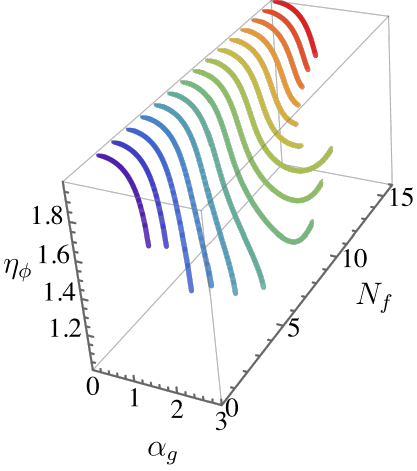}\hspace{1cm}
	\includegraphics[width=.27\columnwidth]{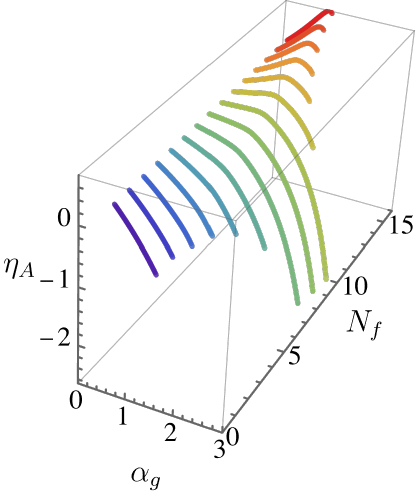}
	\caption{From left to right, $\eta_\psi$, $\eta_\phi$, and $\eta_A$ evaluated along the integrated flows for different $N_f$ and $\alpha_g$ at fixed $N_c=3$.}
	\label{fig:etas}
\end{figure}
In \cref{fig:etas}, we show the anomalous dimensions evaluated along the integrated RG flows for different $N_f$ and $\alpha_g$. The fermion anomalous dimension $\eta_\psi$ grows with the gauge coupling strength, and the symmetric gap condition discussed in \cref{sec:criticalregion} is appreciably satisfied. Furthermore, the meson anomalous dimension increases towards its canonical value, indicating that the mesonic fields are no longer the appropriate degrees of freedom in this region. Finally, $\eta_A$ displays roots which map onto those of the CBZ fixed point.

\subsection{Four-fermion couplings}
\label{app:4Fermidressings}
We now consider the flow of the four-fermion couplings in \eqref{eq:eff4Fermi}. These are extracted from the flow of the four-point functions $\Gamma_k^{(\bar\psi\psi\bar\psi\psi)}$, projected onto the respective tensor structure ${\cal T}_i$, as given in \eqref{eq:4Fermiflow}. Here we employ normalised projectors $\mathbbm{P}^{(\bar\psi\psi\bar\psi\psi)}_i$, as in \cite{Ihssen:2024miv,Gehring:2015vja}, which isolate the individual channels and can be constructed using existing packages~\cite{Braun:2025gvq}.  The diagrammatic flows read,
\begin{subequations}
	\label{eq:flows4F}
	\begin{align}
		\overline{{\rm Flow}}^{(\bar\psi\psi\bar\psi\psi)}_{({\rm S-P})}
		&=\frac{1}{16\pi^2}\Bigg[
		-9\left(1-\frac{\eta_A}{9}-\frac{\eta_\psi}{15}\right)
		\frac{g^4(3N_c^2-8)}{16N_c}
		+\left(1-\frac{\eta_A}{12}-\frac{\eta_\psi}{10}\right)
		\frac{12g^2\bar\lambda_+}{N_f}
		\nonumber\\
		&\qquad
		+\frac{h^2}{30(\lambda_1+1)^2N_cN_f}
		\left(
		2N_c(\bar\lambda_--\bar\lambda_+)
		-\bar\lambda_{\rm adj}(1-N_cN_f)
		\right)
		\nonumber\\
		&\qquad\qquad\times
		\left(
		10-(5\eta_\phi+\eta_\psi)
		+30\left(1-\frac{\eta_\psi}{5}\right)\lambda_1
		\right)
		\Bigg]\,,
	\end{align}
	\begin{align}
		\overline{{\rm Flow}}^{(\bar\psi\psi\bar\psi\psi)}_{({\rm V-A})}
		&=\frac{1}{16\pi^2}\Bigg[
		-9\left(1-\frac{\eta_A}{9}-\frac{\eta_\psi}{15}\right)
		\frac{g^4(N_c^2-1)N_f}{8N_c^2}
		+3\left(1-\frac{\eta_A}{12}-\frac{\eta_\psi}{10}\right)
		\frac{g^2\bar\lambda_{\rm VA}(N_c^2-1)}{N_c^2}
		\nonumber\\[.75ex]
		&\qquad
		+\frac{h^4N_f^2}{8(1+\lambda_1)^3N_c}
		\left(
		(3+\lambda_1)-\frac{\eta_\phi}{3}
		-(1+\lambda_1)\frac{\eta_\psi}{5}
		\right)
		\nonumber\\[.75ex]
		&\qquad
		-\frac{h^2\bar\lambda_+N_f}{(\lambda_1+1)^2}
		\left(
		(2+\lambda_1)-\frac{\eta_\phi}{6}
		-\frac{\eta_\psi}{5}(1+\lambda_1)
		\right)
		\nonumber\\[.75ex]
		&\qquad
		+\frac{1}{2}\left(1-\frac{\eta_\psi}{5}\right)
		\Bigg[
		\frac{\bar\lambda_-^2(1-N_cN_f)}{N_f}
		+\frac{\bar\lambda_-\bar\lambda_{\rm adj}(1-N_c^2)}{N_cN_f}
		-\frac{\bar\lambda_{\rm adj}^2(1-N_c^2)}{N_c^2N_f}
		-\bar\lambda_+^2N_c
		\Bigg]
		\Bigg]\,,
	\end{align}
	\begin{align}
		\overline{{\rm Flow}}^{(\bar\psi\psi\bar\psi\psi)}_{({\rm V+A})}
		&=\frac{1}{16\pi^2}\Bigg[
		9\left(1-\frac{\eta_A}{9}-\frac{\eta_\psi}{15}\right)
		\frac{g^4(N_c^2+4)N_f}{32N_c^2}
		+6\left(1-\frac{\eta_A}{12}-\frac{\eta_\psi}{10}\right)
		\frac{g^2\bar\lambda_+}{N_c}
		\nonumber\\[.75ex]
		&\qquad
		+\frac{h^4N_f}{8(1+\lambda_1)^3}
		\left(
		(3+\lambda_1)-\frac{\eta_\phi}{3}
		-(1+\lambda_1)\frac{\eta_\psi}{5}
		\right)
		\nonumber\\[.75ex]
		&\qquad
		-\frac{h^2
			\left(
			2\bar\lambda_-N_cN_f
			+\bar\lambda_{\rm adj}(N_c-N_f)
			\right)}
		{2(\lambda_1+1)^2N_c}
		\left(
		(2+\lambda_1)-\frac{\eta_\phi}{6}
		+\frac{\eta_\psi}{5}(1+\lambda_1)
		\right)
		\nonumber\\[.75ex]
		&\qquad
		+\left(1-\frac{\eta_\psi}{5}\right)
		\left[
		\frac{4\bar\lambda_-\bar\lambda_+(N_cN_f+1)}{N_f}
		+\frac{2\bar\lambda_{\rm adj}\bar\lambda_+(N_c^2-1)}{N_cN_f}
		+\frac{6\bar\lambda_+^2}{N_f}
		\right]
		\Bigg]\,,
	\end{align}
	\begin{align}
		\overline{{\rm Flow}}^{(\bar\psi\psi\bar\psi\psi)}_{({\rm V-A})_{\rm adj}}
		&=\frac{1}{16\pi^2}\Bigg[
		9\left(1-\frac{\eta_A}{9}-\frac{\eta_\psi}{15}\right)
		\frac{g^4(N_c^2-8)N_f}{16N_c}
		-12\left(1-\frac{\eta_A}{12}-\frac{\eta_\psi}{10}\right)
		\frac{g^2(\bar\lambda_{\rm adj}-\bar\lambda_-N_c)}{N_c}
		\nonumber\\[.75ex]
		&\qquad
		+\frac{h^4N_f^2}{4(1+\lambda_1)^3}
		\left(
		(3+\lambda_1)-\frac{\eta_\phi}{3}
		-(1+\lambda_1)\frac{\eta_\psi}{5}
		\right)
		\nonumber\\[.75ex]
		&\qquad
		+\left(1-\frac{\eta_\psi}{5}\right)
		\left[
		\frac{\bar\lambda_{\rm adj}^2
			(N_c^2+N_cN_f+4)}{N_cN_f}
		-\frac{8\bar\lambda_-\bar\lambda_{\rm adj}}{N_f}
		\right]
		\Bigg]\,.
	\end{align}
\end{subequations}
As dynamical bosonisation is only performed in the (S--P) channel the respective four-fermion couplings are absent and instead we have the Yukawa parameters. Simultaneously, ${\cal A}_j=0$ for $j=\{\pm,\,{\rm adj}\}$.  The flows in \eqref{eq:flows4F} have three types of contributions: pure gauge box diagrams, mixed four-fermion--gauge diagrams, and pure Yukawa or four-fermion diagrams. The first propagate the gauge relevant deformation into the higher fermionic operators and provide the initial contribution that generates a non-vanishing tower of fermionic operators.

\subsection{Yukawa coupling}
\label{app:Yukawa}
The Yukawa coupling $h$ appears as a consequence of bosonisation and carries part of the information contained in the higher-dimensional interactions. Its flow can be extracted from the fermion two-point function in the scalar channel, as in \cite{Ihssen:2024miv,Goertz:2024dnz,Fu:2019hdw}, or from any fermion-meson vertex,
\begin{align}
	\partial_t h_i
	&=
	\left(\frac{1}{2}\eta_{\phi_i}+\eta_\psi\right)h
	-\bar m_{\phi_i}^2\,\dot{\cal A}_i
	+\frac{
		{\rm Tr}
		\left[
		{\cal P}^{(\bar\psi\psi\phi_i)}
		\partial_t\Gamma_k^{(\bar\psi\psi\phi_i)}
		\right]_{p=0}
	}{
		Z_\psi Z_{\phi_i}^{1/2}
		{\rm Tr}
		\left[
		{\cal P}^{(\bar\psi\psi\phi_i)}
		{\cal P}^{(\bar\psi\psi\phi_i)}
		\right]
	}\,,
\end{align}
with $i=\{\sigma,\pi,a,\eta\}$ and \eqref{eq:dynbos_cond}. In the broken phase, the different scalar modes become distinguishable and the Yukawa couplings for each mode therefore differ. In this work, we have prepared the bosonisation for accessing the broken phase but focus the analysis on the symmetric phase, where all modes are degenerate. Furthermore, the rightmost term contains the diagrammatic flow, which in the symmetric phase reads
\begin{align}
	\frac{
		{\rm Tr}
		\left[
		{\cal P}^{(\bar\psi\psi\phi_i)}
		\partial_t\Gamma_k^{(\bar\psi\psi\phi_i)}
		\right]_{p=0}
	}{
		Z_\psi Z_{\phi_i}^{1/2}
		{\rm Tr}
		\left[
		{\cal P}^{(\bar\psi\psi\phi_i)}
		{\cal P}^{(\bar\psi\psi\phi_i)}
		\right]
	}
	=
	\frac{h}{16\pi^2}
	\left[
	\frac{3g^2(1-N_c^2)}{N_c}
	\left(1-\frac{\eta_A}{12}-\frac{\eta_\psi}{10}\right)
	+\frac{8\bar\lambda_+}{N_f}
	\left(1-\frac{\eta_\psi}{5}\right)
	\right]\,,
\end{align}
and is common to all mesonic modes and to the scalar-channel fermion two-point function.

\subsection{Flow of the effective potential}
\label{app:potential}
We now tackle the flow of the effective potential including the $\tau$ invariant and for this we directly follow the derivation in \cite{Jungnickel:1995fp} where this was first derived. The flow of the dimensionless effective potential $u(\bar\rho,\bar\tau)$, expressed in terms of the dimensionless renormalised fields $\bar\rho$ and $\bar\tau$, reads
\begin{align}
	\partial_t u(\bar\rho,\bar\tau)
	&=
	\left.\partial_t\right|_{\rho,\tau}
	\frac{V(\rho,\tau)}{k^4}
	-4u(\bar\rho,\bar\tau)
	+(2-\eta_\phi)\bar\rho\,\partial_{\bar\rho}u(\bar\rho,\bar\tau)
	+(4-2\eta_\phi)\bar\tau\,\partial_{\bar\tau}u(\bar\rho,\bar\tau)\,,
	\label{eq:flowU}
\end{align}
where the last two terms encode the canonical and anomalous scaling of the two invariants. The first term corresponds to the flow of the effective potential at fixed fields and is obtained from the diagrammatic expression
\begin{align}
	\left.\partial_t\right|_{\rho,\tau}V(\rho,\tau)
	&=
	\frac{k^4}{32\pi^2}
	\left(1-\frac{\eta_\phi}{6}\right)
	\,{\rm Tr}
	\left[
	\frac{1}{\mathbbm{1}+\bar M_{\phi,ab}^2}
	\right]_{\rm bos}
	-\frac{k^4N_c}{8\pi^2}
	\,{\rm Tr}
	\left[
	\frac{1}{\mathbbm{1}+\bar m_{\psi,ab}^2}
	\right]_{\rm fer}\,,
	\label{eq:flowVfull}
\end{align}
with the first contribution corresponding to the mesonic modes and the second to the fermions. We have omitted the gauge and ghost contributions, as they do not enter any of the flows relevant here. In \eqref{eq:flowVfull}, we have performed the internal momentum integral and traced over the colour and Lorentz structures, while the flavour matrix remains to be traced, which requires particular care.

Let us first consider the meson-loop contribution, whose two-point function in flavour space reads
\begin{align}
	\Gamma_k^{(\phi_a\phi_b)}\big|_{\phi=\phi_0}
	&=
	\frac{\delta^2\Gamma_k}{\delta\phi_a\delta\phi_b}
	=
	Z_\phi\,p^2\delta_f^{ab}+M_{\phi,ab}^2\,,
	\label{eq:Gamma2LPA}
\end{align}
and the flavour dependence is resolved by diagonalising the mass matrix
$M_{\phi,ab}^2=\partial^2V(\rho,\tau)/\partial\phi_a\partial\phi_b$, i.e.\ the Hessian, at an appropriate background. Since the potential depends on the fields only through the chiral invariants, applying the chain rule gives 
\begin{align}
	M_{\phi,ab}^2
	&=
	V_\rho\,\rho_{ab}
	+V_\tau\,\tau_{ab}
	+V_{\rho\rho}\,\rho_a\rho_b
	+V_{\rho\tau}
	(\rho_a\tau_b+\tau_a\rho_b)
	+V_{\tau\tau}\,\tau_a\tau_b\,,
	\label{eq:Hessian}
\end{align}
where we use the shorthand $\rho_a=\partial\rho/\partial\phi_a$,
$\rho_{ab}=\partial^2\rho/\partial\phi_a\partial\phi_b$, and likewise for $\tau$. Since $\rho={\rm Tr}[\Sigma^\dagger\Sigma]$ is quadratic in the fields, one has $\rho_{ab}=\delta_{ab}$ for all field components, so $V_\rho$ contributes universally to all modes. The first derivatives $\rho_a$ are nonzero only for field components aligned with the background.

We now exploit the residual $\left[U(1)\right]^{N_f}$ symmetry of the diagonal background to block-diagonalise the mass matrix, as in \cite{Jungnickel:1995fp}. The off-diagonal field components $\phi_{ab}$ with $a\neq b$ carry definite $U(1)_a$ charge $+1$ and $U(1)_b$ charge $-1$ and therefore cannot mix with components of different charge. Moreover, $\phi_{ab}$ and $\phi_{ba}$ mix with each other but with no other components, giving a $2\times2$ block that further factorises into symmetric $(+)$ and antisymmetric $(-)$ combinations, each with a single eigenvalue.

Furthermore, the field can be also decomposed as $\phi=\phi_R+{\rm i}\phi_I$, where the scalar ($R$) and pseudoscalar ($I$) components decouple because $\partial^2\tau/\partial\phi_R\partial\phi_I=0$ at any real diagonal background. The full $2N_f^2\times2N_f^2$ mass matrix therefore decomposes into independent blocks that can be diagonalised separately.

\paragraph{Non-degenerate background and projection.}
To extract the flow of $V_\tau$, one must evaluate the flow equation at a non-degenerate background $\phi_0$ with $\tau\neq0$, since $\tau=0$ at any $U(N_f)_V$-symmetric configuration and $\partial_\tau(\partial_tV)|_{\tau=0}$ is therefore not accessible at the degenerate background. This is essential for correctly accounting for the $\tau$-dependent corrections even in the symmetric regime. As in \cite{Jungnickel:1995fp}, we use the minimal non-degenerate background
\begin{align}
	\varphi_1^2
	&=
	\frac{\rho}{N_f}
	+\sqrt{\frac{\tau}{N_f(N_f-1)}}\,,
	&
	\varphi_{N_f}^2
	&=
	\frac{\rho}{N_f}
	-\sqrt{\frac{(N_f-1)\tau}{N_f}}\,,
	&
	\varphi_1&=\varphi_2=\ldots=\varphi_{N_f-1}\,,
	\label{eq:minimalbackgroung}
\end{align}
which realises both $\rho$ and $\tau$ with only two distinct field values, exploiting the residual $O(N_f-1)$ symmetry of the background to minimise the number of independent mass eigenvalues. We consider a real diagonal background $\phi_0={\rm diag}(\varphi_1,\ldots,\varphi_{N_f})$, where only the diagonal scalar directions are nonzero. This choice is purely technical and has no physical implication. However, it is the most convenient because it provides the simplest non-trivial background that allows us to distinguish $\rho$ from $\tau$.

\paragraph{Pseudoscalar sector.}
For the pseudoscalar fields $\phi_{I,ab}$, the first derivatives $\tau_a^{(I)}=0$ at any real diagonal background, because $\tau$ as defined in \eqref{eq:rhotau} contributes to $\tau_{ab}^{(I)}$ only through combinations of commutators of generators that yield antisymmetric contributions vanishing under the trace. As a result, the pseudoscalar Hessian contains no $V_{\rho\rho}$, $V_{\rho\tau}$, or $V_{\tau\tau}$ terms. The $N_f$ diagonal pseudoscalar modes have eigenvalues
	\begin{align}
		\bar M_{I,a}^2
		&=
		\frac{2V_\tau}{N_f}
		\left(N_f\varphi_a^2-\rho\right)
		+V_\rho\,,
		\label{eq:MIa}
	\end{align}
	and the $N_f(N_f-1)/2$ pairs of off-diagonal pseudoscalar modes have eigenvalues
	\begin{align}
		\left(\bar M_{I,ab}^-\right)^2
		&=
		\frac{2V_\tau}{N_f}
		\left[
		N_f(\varphi_a^2+\varphi_b^2+\varphi_a\varphi_b)-\rho
		\right]
		+V_\rho\,,
		\label{eq:MIabm}
		\\
		\left(\bar M_{I,ab}^+\right)^2
		&=
		\frac{2V_\tau}{N_f}
		\left[
		N_f(\varphi_a^2+\varphi_b^2-\varphi_a\varphi_b)-\rho
		\right]
		+V_\rho\,.
		\label{eq:MIabp}
	\end{align}
	At the degenerate background $\varphi_a^2=\rho/N_f$, all three pseudoscalar eigenvalues reduce to $V_\rho$, which is the pion (Goldstone) mass squared at the minimum $V_\rho(\rho_0)=0$.
	
	\paragraph{Scalar off-diagonal sector.}
	For the off-diagonal scalar modes $\phi_{R,ab}$ with $a\neq b$, the analysis is identical to the pseudoscalar case, since at the diagonal background the first derivative of $\tau$ with respect to $\phi_{R,ab}$ produces the same bilinear structure. The eigenvalues are
	\begin{align}
		(\bar M_{R,ab}^+)^2
		&=(\bar M_{I,ab}^-)^2,
		&
		(\bar M_{R,ab}^-)^2
		&=(\bar M_{I,ab}^+)^2\,,
		\label{eq:MRab}
	\end{align}
	so the scalar and pseudoscalar off-diagonal modes occur in degenerate pairs with interchanged $\pm$ labels.
	
\paragraph{Scalar diagonal sector.}
The $N_f\times N_f$ block of diagonal scalar fields $\phi_{R,aa}$ is the most involved sector, as the $V_{\rho\rho}$, $V_{\rho\tau}$, and $V_{\tau\tau}$ terms contribute,
\begin{align}
		\tilde M_{R,ac}^2
		&=
		\left[
		V_\rho
		+\frac{2V_\tau}{N_f}
		\left(3N_f\varphi_a^2-\rho\right)
		\right]\delta_{ac}
		+2\varphi_a\varphi_c
		\Bigg[
		V_{\rho\rho}
		+\frac{2V_{\rho\tau}}{N_f}
		\left(N_f(\varphi_a^2+\varphi_c^2)-2\rho\right)
		\nonumber\\
		&\qquad
		+\frac{4V_{\tau\tau}}{N_f^2}
		(N_f\varphi_a^2-\rho)(N_f\varphi_c^2-\rho)
		-\frac{2V_\tau}{N_f}
		\Bigg]\,.
		\label{eq:MRac}
\end{align}
This matrix has no simple closed-form eigenvalues at a general background but can nonetheless be diagonalised by defining
\begin{subequations}
		\begin{align}
			A
			&=
			\tilde M_{R,11}^2
			+(N_f-2)\tilde M_{R,12}^2
			\big|_{\varphi_2\to\varphi_1}\,,
			&
			B&=\tilde M_{R,N_fN_f}^2\,,
			&
			C&=\sqrt{N_f-1}\,\tilde M_{R,1N_f}^2\,,
			\label{eq:ABCblock}
		\end{align}
		with eigenvalues
		\begin{align}
			\lambda_\pm
			&=
			\frac{A+B}{2}
			\pm
			\sqrt{
				\left(\frac{A-B}{2}\right)^2+C^2
			}\,,
			\label{eq:lambdapm}
		\end{align}
	\end{subequations}
which reduce to $\lambda_+=\bar m_\sigma^2$ and $\lambda_-=\bar m_a^2$ at the degenerate background $\varphi_a^2=\rho/N_f$ and $\tau=0$, as required.

\begin{figure}[t!]
	\centering
	\includegraphics[width=.65\columnwidth]{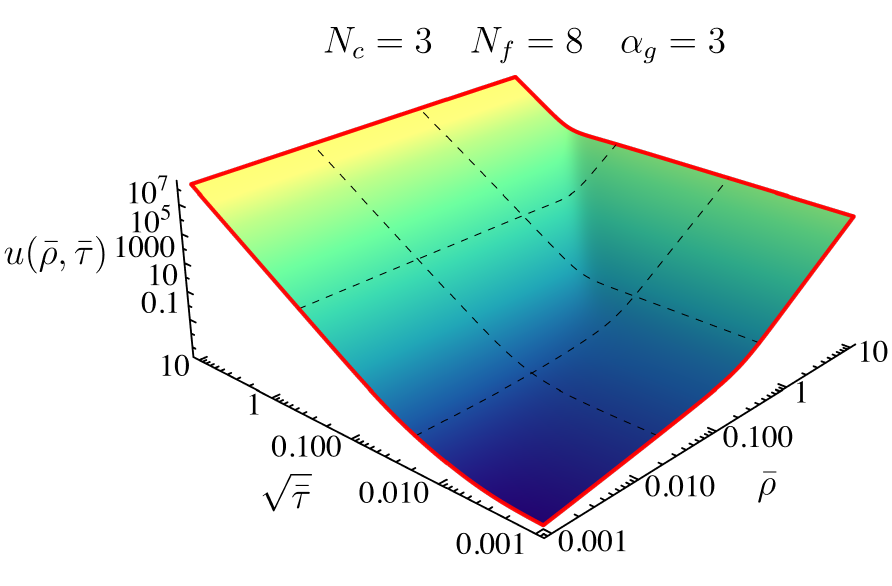}
	\caption{Integrated effective mesonic potential for a theory with $N_c=3$ and $N_f=8$ at strong coupling $\alpha_g=3$.}
	\label{fig:effectiveVNf8}
\end{figure}

The total bosonic threshold sum in \eqref{eq:flowVfull} over all $2N_f^2$ modes is then
\begin{align}
{\rm Tr}\left[\frac{1}{\mathbbm{1}+\bar M_{\phi,ab}^2}\right]_{\rm bos}&=\frac{N_f-2}{1+(\bar M_{Rab}^+)^2|_{\varphi_1,\varphi_1}}+\frac{1}{1+\lambda_+}+\frac{1}{1+\lambda_-}\notag\\
&\quad+\frac{N_f-1}{1+\bar M_{Ia}^2|_{\varphi_1}}+\frac{1}{1+\bar M_{Ia}^2|_{\varphi_{N_f}}}\notag\\
&\quad+\frac{(N_f-1)(N_f-2)}{2}\left[\frac{1}{1+(\bar M_{Rab}^+)^2|_{\varphi_1,\varphi_1}}+\frac{1}{1+(\bar M_{Rab}^-)^2|_{\varphi_1,\varphi_1}}\right]\notag\\
&\quad+(N_f-1)\left[\frac{1}{1+(\bar M_{Rab}^+)^2|_{\varphi_1,\varphi_{N_f}}}+\frac{1}{1+(\bar M_{Rab}^-)^2|_{\varphi_1,\varphi_{N_f}}}\right]\notag\\
&\quad+\frac{(N_f-1)(N_f-2)}{2}\left[\frac{1}{1+(\bar M_{Iab}^+)^2|_{\varphi_1,\varphi_1}}+\frac{1}{1+(\bar M_{Iab}^-)^2|_{\varphi_1,\varphi_1}}\right]\notag\\
&\quad+(N_f-1)\left[\frac{1}{1+(\bar M_{Iab}^+)^2|_{\varphi_1,\varphi_{N_f}}}+\frac{1}{1+(\bar M_{Iab}^-)^2|_{\varphi_1,\varphi_{N_f}}}\right]\,.\label{eq:flowVtraceboson}
\end{align}

\paragraph{Fermion sector.}
Returning to the fermion loop in \eqref{eq:flowVfull} and using the minimal non-degenerate background in \eqref{eq:minimalbackgroung}, the flavour trace is straightforwardly evaluated as
\begin{align}
	{\rm Tr}
	\left[
	\frac{1}{\mathbbm{1}+\bar m_{\psi,ab}^2}
	\right]_{\rm fer}
	=
	\frac{N_f-1}{1+h^2\bar\varphi_1^2}
	+\frac{1}{1+h^2\bar\varphi_{N_f}^2}\,.
	\label{eq:Ffer}
\end{align}
This yields two distinct masses, which coincide on the degenerate background but still allow the $\tau$ dependence to be disentangled in the symmetric phase.

\printbibliography
	
\end{document}